\documentclass{article}

\usepackage{arxiv}

\usepackage[utf8]{inputenc} % allow utf-8 input
\usepackage[T1]{fontenc}    % use 8-bit T1 fonts
\usepackage{hyperref}       % hyperlinks
\usepackage{url}            % simple URL typesetting
\usepackage{booktabs}       % professional-quality tables
\usepackage{amsfonts}       % blackboard math symbols
\usepackage{nicefrac}       % compact symbols for 1/2, etc.
\usepackage{microtype}      % microtypography
\usepackage{lipsum}
\usepackage{url}
\usepackage{breakurl}

\usepackage{hhline}
\usepackage{longtable}
\usepackage{supertabular}
\usepackage{multirow}
\usepackage{tabu}
\usepackage{rotating, graphicx}
\usepackage [autostyle, english = american]{csquotes}
\MakeOuterQuote{"}

\usepackage{xcolor}
\usepackage{color, colortbl}
\usepackage[
singlelinecheck=false % <-- important
]{caption}
\title{Architecture and Security of SCADA Systems \\ : A Review}

\author{
  Geeta Yadav \\
  School of Information Technology\\
  IIT Delhi, India\\
  \texttt{geeta@cse.iitd.ac.in} \\
   \And
 Kolin Paul \\
  Department of Computer Science\\
  IIT Delhi, India\\
  \texttt{kolin@cse.iitd.ac.in} \\
}

\begin{document}
\maketitle

\begin{abstract}
Pipeline bursting, production lines shut down, frenzy traffic, trains confrontation, nuclear reactor shut down, disrupted electric supply, interrupted oxygen supply in ICU - these catastrophic events could result because of an erroneous SCADA system/ Industrial Control System(ICS).  SCADA systems have become an essential part of automated control and monitoring of many of the Critical Infrastructures (CI).  Modern SCADA systems have evolved from standalone systems into sophisticated complex, open systems, connected to the Internet. This geographically distributed modern SCADA system is vulnerable to threats and cyber attacks. In this paper, we first review the SCADA system architectures that have been proposed/implemented followed by attacks on such systems to understand and highlight the evolving security needs for SCADA systems. A short investigation of the current state of intrusion detection techniques in SCADA systems is done
%     . Various  risk assessment proposals are discussed and categorised
    , followed by a brief study of testbeds for SCADA systems. The cloud and Internet of things (IoT) based SCADA systems are studied by analysing the architecture of modern SCADA systems.  This review paper ends by highlighting the critical research problems that need to be resolved to close the gaps in the security of SCADA systems.
\end{abstract}

% keywords can be removed
\section{Introduction}\label{introduction}
\begin{figure}[h!]
	\centering
	\includegraphics[width=3.5in]{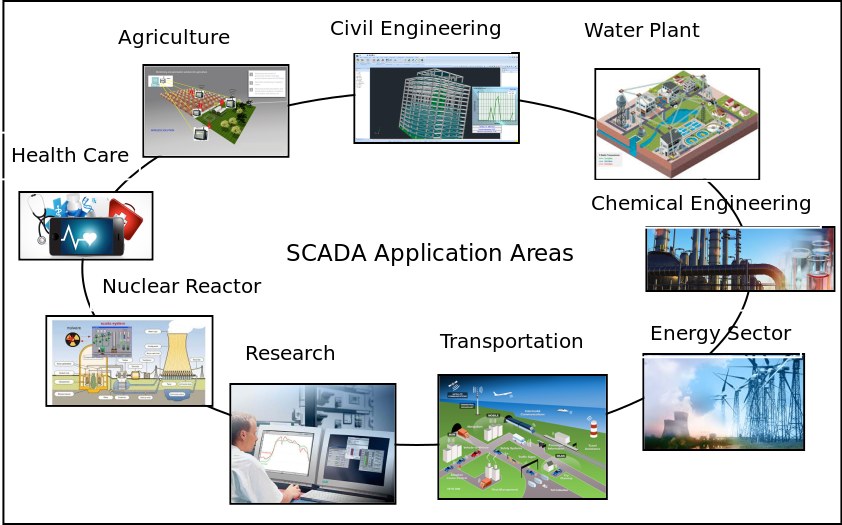}
	\caption{SCADA Application Areas }
	\label{fig:applicationarea}
\end{figure}

Critical Infrastructures (CI) are often described as the infrastructures which provide essential services and serves as the foundation for any nation's security, economy, and healthcare systems. Cyber-Physical Systems (CPS)/ Internet of Things (IoT), are supplementing traditional
CI with data-rich operations. The list of sectors under critical infrastructure varies from country to country. \textcolor{black}{It generally includes sectors like agriculture,  healthcare, nuclear reactor, transportation, energy sector, civil and chemical engineering, water plants, research etc. as depicted in Fig. \ref{fig:applicationarea}}. Supervisory Control and Data Acquisition (SCADA) systems, an  Industrial Control Systems(ICS), have a pivotal role in managing and controlling of the CI. SCADA systems control and monitor geographically distributed assets. Historically, SCADA frameworks were limited to power transmission, gas conveyance, and water appropriation control frameworks. Advancements in technology have led to SCADA being deployed in steel making, chemical processing industries, telecommunications, experimental and manufacturing facilities \cite{Nasr2015}. With Industries 4.0 / Industrial Internet of Things (IIoT) evolution, modern SCADA systems have adopted CPS/ IoT, cloud technology, big data analytics, Artificial intelligence (AI) and machine learning. Integration of these technologies has significantly improved interoperability, ease the maintenance and decreased the infrastructure cost. Therefore, leading to a near real-time environment. \par 
SCADA systems improve the efficiency of the operation of the industrial critical system as well as provide better protection to the utilised equipment. Moreover, it improves the productivity of the personnel. SCADA frameworks give valid identification and prompt alert warning to the observing stations by using an attested monitoring stage, advanced communications, and state-of-the-art sensors.
SCADA systems were designed to work in a standalone way and relied on air-gapped networks and proprietary protocols for securing the system. Therefore, initial designs of SCADA never incorporated security features \cite{Rezai2016,PAPA201396}. However in recent years, due to the expansion of business and need of central monitoring of distributed software, SCADA systems have evolved into sophisticated, complex open systems, connected to the Internet using advanced technology. Associating SCADA system to the web has helped numerous SCADA systems to work from topographically inaccessible areas. However, this has lead the SCADA system more vulnerable for attackers to target from anywhere in the world  \cite{Miller:2012:SSC:2380790.2380805}. \par The modernisation of the SCADA system, standardisation of communication protocols and growing interconnectivity have drastically increased the cyber-attacks on SCADA system over the years. These type of attacks are becoming more sophisticated to commit the more traditional cyber espionage and sabotage in addition to cyber crimes. The smooth and genuine operation of SCADA framework is one of the key concern for the enterprises, because the outcome of break down of SCADA system may range from financial misfortune to natural harm to loss of human life \cite{Cherdantseva2016}.
A cyber-attack on a nuclear plant will have a global impact. Moreover, the security spillage in small networks can lead to a loss of services and financial loss to the utility company. \par
\textcolor{black}{Many international institutes e.g. IEEE, Centre for the Protection of National Infrastructure (CPNI), 
American Gas Association (AGA), Industrial Automation and Control System Security (ISA), North American Electric Reliability Corporation (NERC) and National Institute of Standards and Technology (NIST) etc. publish guidelines frequently for secure SCADA implementation \cite{Sommestad2010a}.}
% \textcolor{black}{Therefore, many international institutes e.g. IEEE, Centre for the Protection of National Infrastructure (CPNI) \cite{CPNI}, 
% American Gas Association (AGA) \cite{aga}, Industrial Automation and Control System Security (ISA) \cite{ISA}, North American Electric Reliability Corporation (NERC)\cite{NERC} and National Institute of Standards and Technology (NIST) \cite{nist} etc. publish guidelines frequently for secure SCADA implementation \cite{Sommestad2010a}.}
% This paper reviews modern SCADA system architecture, the potential vulnerabilities, attacks on the SCADA system, intrusion detection systems, and testbeds for the SCADA system.
% risk assessment models implemented in a SCADA system.

\par

% \subsection{Organization}
% We target to present systematically existing knowledge and identify open problems. Section \ref{taxonomy} gives an overview of taxonomy used for research of architecture and security of SCADA systems.
% Section \ref{architecture} gives an outline of the architecture of SCADA systems. Section \ref{Taxonomy of attacks} discusses a consolidated history of the reported attacks on the SCADA systems. This section also covers the quantitative assessments of the attacks and their impact on systems and environments. 
%  Section \ref{Cause of vulnerability} discusses aspects of SCADA systems security vulnerabilities, and threats.
% Section \ref{Intrusion Detection} covers intrusion detection system for SCADA systems. 
% % Section \ref{Risk Assessment} present a taxonomy of risk assessments methods for SCADA systems.
% Section \ref{Testbeds} discusses the taxonomy of government and industrial testbeds. Section \ref{Recent advances in SCADA} covers IoT-Cloud based SCADA systems. The review ends with Section \ref{Future Research Directions} where we identify the future scope of research in SCADA systems.
\begin{figure*}[b!]
	\centering
	\includegraphics[width = 0.8\textwidth]{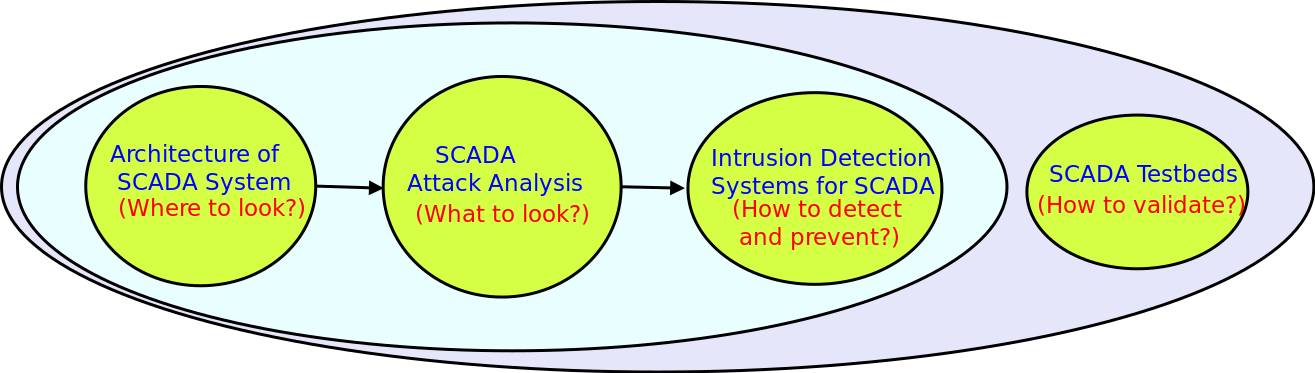}
	\caption[]{Relation between different dimensions of SCADA security}
	\label{fig:dimens_connect}
\end{figure*}

\begin{figure*}[b!]
	\centering
	\includegraphics[width = 1\textwidth, height =9cm]{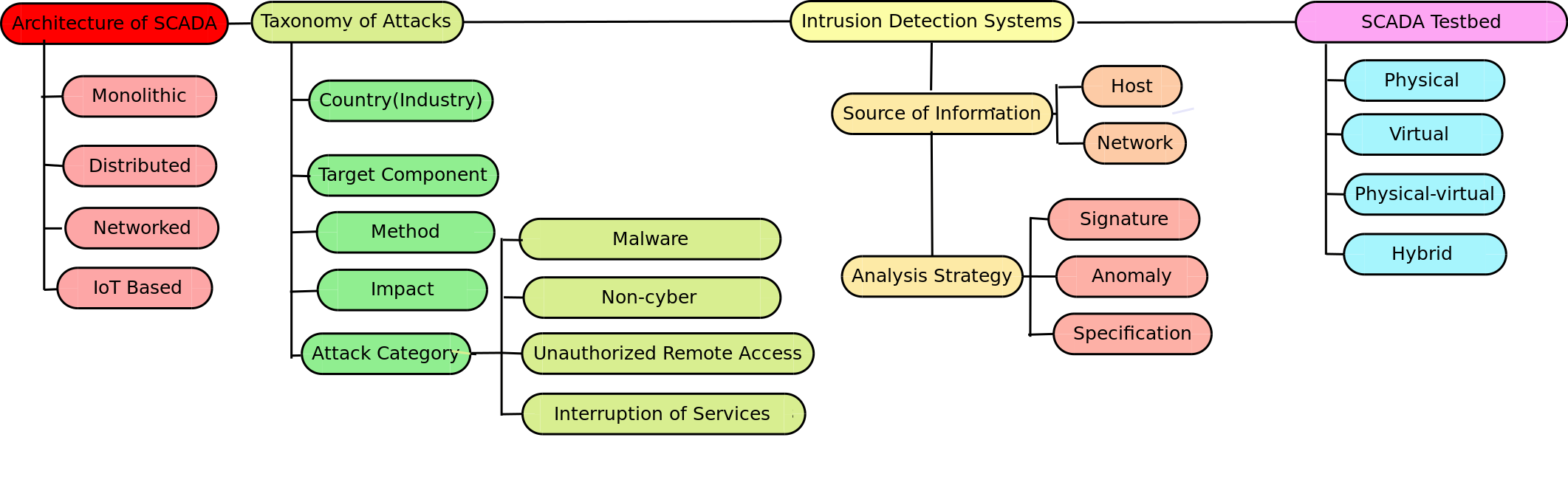}
	\caption[]{Taxonomy of Research}
	\label{fig:Taxonomy}
\end{figure*}
\subsection{Scope}
\textcolor{black}{To the best of our knowledge, this is the first work which discusses and seeks to interconnect the various aspects of SCADA systems ranging from architecture, vulnerabilities and attacks, Intrusion Detection Systems \& techniques and the testbeds as shown in Fig, \ref{fig:dimens_connect}. This allows for a more complete and holistic view of SCADA system security. We seek to answer the question "where to look for security vulnerabilities" by explaining the interconnection between SCADA architecture, the communication protocols. The linking between the communication and the vulnerabilities in the systems help answer "what to look for?". The mutual dependencies of the protocols, existing intrusion detection and prevention mechanisms and the vulnerabilities should be considered for detection and prevention of security issues. The lessons learned and the hardening techniques developed can only be deployed on the SCADA systems post rigorous validation on testbeds.}
\textcolor{black}{The surveys published so far discuss and detail only one aspect of the SCADA security and thus fail to the show the interconnections between various dimension that is essential to design security mechanisms for the complex IIoT systems of the future.}
\textcolor{black}{Thus, the motive of this review is to link the different aspects of SCADA security while considering their known loopholes.
}

% \textcolor{black}{To the best of our knowledge, this is the first work which discusses the architecture of SCADA systems in depth (to identify the scope of vulnerable components). It demonstrates the advancement of SCADA frameworks in various dimensions of security ranging from vulnerabilities, attack databases (to relate the real-time attacks to vulnerable components), intrusion detection systems (tactics for detecting and preventing attacks ) and the various SCADA testbeds (to validate the security techniques feasibility and performance) as shown in Fig, \ref{fig:dimens_connect}.} 
% %This is the first work which discusses SCADA systems in different dimensions in our knowledge. 
% However, few surveys have been published considering one dimension. These survey lack in the interconnection between different dimension. 
% % Initially, by analysing attacks on that system and testing, the vulnerabilities present in the system are highlighted.  To secure the SCADA system, an IDS needs to work well, be accurate, needs to be tested thoroughly on testbeds and it should be secure itself. Therefore, there is a need for a practical approach rather than theoretical techniques for IDS validation. A framework for assessing risks of the SCADA system increases assurance for the functionality.
% Thus, the motive of this review is to link the different aspects of SCADA security while considering their known loopholes.

\subsection{Review methodology} This section describes the approach taken for selecting the various relevant papers and then classifying their work. 
We checked conference and journal ranking specifically for SCADA cyber-security, Industrial security, and CI. \par  We gathered and assessed related documents from these top conferences and journals. Apart from that, searches were done on IEEE Xplore, ACM, IET journal, and SCOPUS, which lead to excellent coverage of useful publications. Then we categorised papers based on IDS, testbeds,
% risk assessment 
and attacks analysis manually.Next, we reviewed documents, section by section, based on the examination of the title, abstract and full text in case paper provide a novel idea. We then correlated the various work done with the different dimensions of SCADA security, resulting in a corresponding taxonomy.
Related work related to each dimension is dicussed in respective section.
%  \subsection{Related Work}
%  To the best of our knowledge, this is the first work which discusses the architecture of SCADA systems in depth and then demonstrates the advancement of SCADA frameworks in various dimensions of security ranging from vulnerabilities, attack databases, intrusion detection systems and the various SCADA testbeds. 
% %This is the first work which discusses SCADA systems in different dimensions in our knowledge. 
% However, few surveys have been published considering one dimension. These survey lack in the interconnection between different dimension. 
% % Initially, by analysing attacks on that system and testing, the vulnerabilities present in the system are highlighted.  To secure the SCADA system, an IDS needs to work well, be accurate, needs to be tested thoroughly on testbeds and it should be secure itself. Therefore, there is a need for a practical approach rather than theoretical techniques for IDS validation. A framework for assessing risks of the SCADA system increases assurance for the functionality.
% Thus, the motive of this review is to link the different aspects of SCADA security while considering their known loopholes.
%  \par Risk assessment approaches related to SCADA system are reviewed in \cite{Cherdantseva2016} by selecting  24 risk
% assessment methods developed in SCADA system context. Author analysed these approaches in terms of aim of proposed approach, application domain,
% risk management stage, evaluation  and key concept covered. Author suggested that validation of these approaches is a major concern.
\section{Taxonomy} \label{taxonomy}
\textcolor{black}{We propose a taxonomy for studying the architecture and security aspect of SCADA depicted in Fig. \ref{fig:Taxonomy}. First, we discuss SCADA architecture and its components. The SCADA architecture is classified into four generations, i.e. Monolithic, Distributed, Networked, and IoT based fourth generation. Afterwards, we discuss SCADA specific commonly used communication protocols considering their reference architecture, addressing and security state, as explained in detail in Section \ref{architecture}. }
\par \textcolor{black}{ An analysis of attacks on SCADA system is necessary to develop technology for handling new attacks. We report some real-time SCADA attacks to demonstrate the impact of these attacks on a nation. We aim to show the urgent need for securing SCADA systems. Therefore, we have analysed the attacks based on the country (industry) of attack, the target component, the impact of the attack and the type of attack. We have classified the attacks in five categories, i.e. Malware, Non-cyber attack, Unauthorised remote access,  Interruption of services, and Unknown in Section \ref{Taxonomy of attacks}.}
\par \textcolor{black}{Intrusion detection systems (IDSs) are used to detect and prevent these attacks, and recognising vulnerabilities in the systems. We have categorised IDSs based on the source of information and based on analysis strategy. Source of information can be the host or the network. The analysis strategy can be signature-based, specification-based, anomaly detection, or using machine learning-based algorithms, i.e. clustering-based, probabilistic model-based etc. These IDSs are studied considering the threat model (Attacks handled), required input data and technique considering our taxonomy. This analysis helps to link the security measures taken to avoid a particular attack. A detailed analysis of IDSs is done in Section \ref{Intrusion Detection}.}
\par \textcolor{black}{
Most of IDS tools need to be trained and tested on a relevant and validated dataset, which will be unique for each industry and each SCADA system. 
To overcome the lack of validated datasets, researchers are focusing on creating testbeds for data sets. Apart from that, deploying these IDSs on a running SCADA system is a challenging task as these are part of critical infrastructure which cannot bear a shutdown, delay etc. Therefore, the testbed plays an important role. We have classified testbeds into four categories based on their implementation strategies, i.e. physical testbed, virtual testbed, virtual-physical testbed, probabilistic model-based etc., software, and hybrid testbed and we survey their advantages and disadvantages in Section \ref{Testbeds}. Section \ref{Recent advances in SCADA} discusses IoT-Cloud based SCADA systems. The review ends with Section \ref{Future Research Directions} where we identify the future scope of research in SCADA systems.
}

% To analyse the risk of an attack various risk assessment techniques have been proposed over the years. 
% We have studied these techniques by categorising based on risk calculation method in Section \ref{Risk Assessment}.

%   There is three types of risk calculation have been used, i.e. Subjective, Probabilistic Numeric, Non-probabilistic Numeric. Then we categorised variously proposed algorithm based on risk model used, i.e. attack model, Intent model and Twin model.

\section{SCADA System Architecture} \label{architecture}
\textcolor{black}{SCADA framework is an amalgamation of hardware components and software programs where hardware includes a ``Remote Terminal Units (RTU)'', ``Master Terminal Unit (MTU)'', actuators and sensors, and software includes ``Human Machine Interface (HMI)'', a central database (Historian) and other user software \cite{Shahzad2014a}.} These softwares provide a communication interface between hardware and software.

The physical environment is linked to the actuators and sensors which are further connected to RTUs. RTUs gather the information and data from the sensors and forward telemetry data to the MTU for observing and controlling the SCADA framework. We discuss this in greater detail in the next section. 
\subsection{SCADA Components}
\begin{figure}[h!]
	\centering
	\includegraphics[height=10cm]{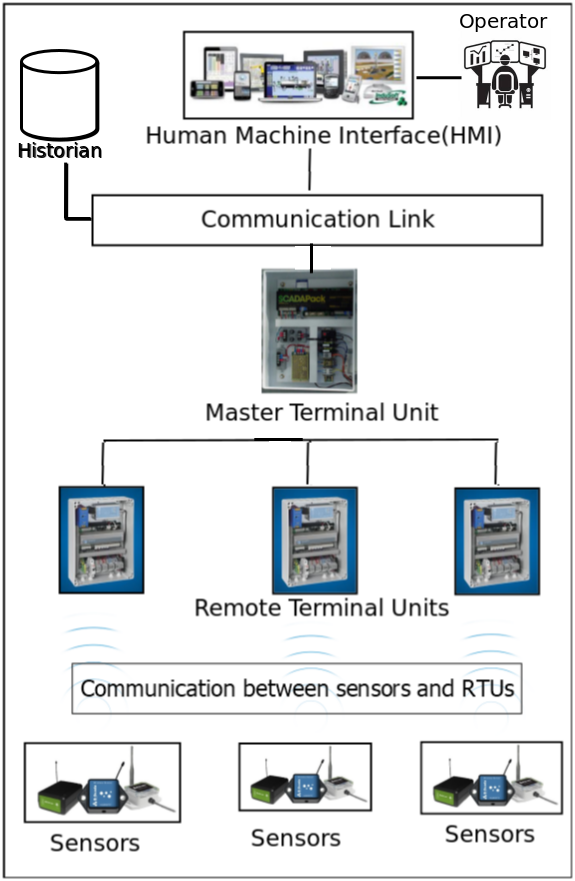}
	\caption{\textcolor{black}{Interrelation of SCADA system components (inspired by \cite{Shahzad2014a})}}
	\label{fig:SCADAarchitecture}
\end{figure}

The interrelation of SCADA system components MTU, RTU, HMI, Historian and SCADA communication links is represented in Fig. \ref{fig:SCADAarchitecture}. 
\textbf{RTU} is responsible for collecting real-time data and information from sensors which are connected to the physical environment using link LAN/WAN. RTUs forward information to MTU. These are additionally in charge of conveying the present status data of physical devices associated with the system.\\
\begin{figure*}[h!]
		\centering
		\begin{minipage}{.5\textwidth}
				\centering
				\includegraphics[width = 2.5in, height = 2.25in]{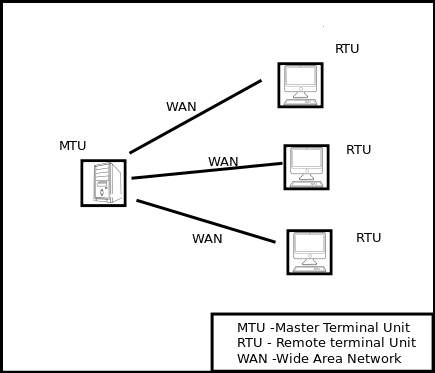}
				\caption{\textcolor{black}{Monolithic SCADA system Architecture}}
				\label{fig:monolithicSCADA}
		\end{minipage}%
		\begin{minipage}{.5\textwidth}
				\centering
				\includegraphics[width = 2.75in, height = 2.25in]{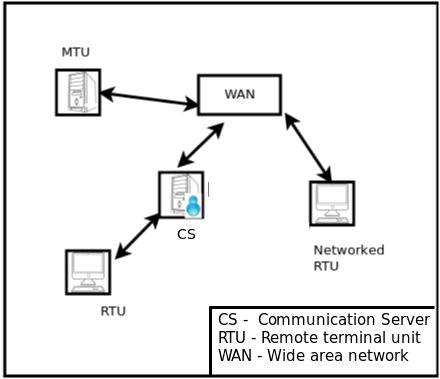}
				\caption{\textcolor{black}{Networked SCADA system Architecture}}
				\label{fig:networkedSCADA}
		\end{minipage}
		\begin{minipage}{1\textwidth}
				\centering
				\includegraphics[width = 3.5in, height = 2.25in]{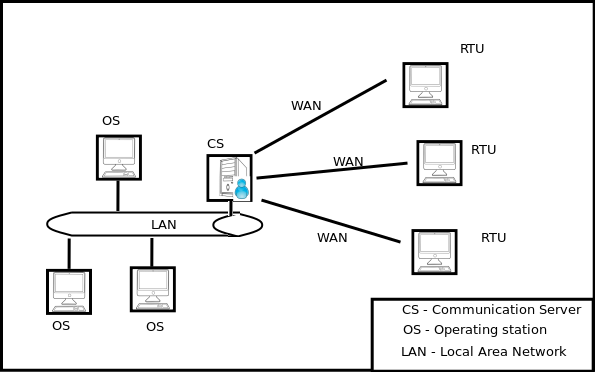}
				\caption{\textcolor{black}{Distributed SCADA system Architecture}}
				\label{fig:distributedSCADA}
		\end{minipage}
		
\end{figure*}
\textbf{MTU} is the central monitoring station. It is in charge of controlling and commanding the RTU machine over communication links. It also responds to messages from RTU and processes and stores them for succeeding communication.\\
\textbf{HMI} provides a communication interface between SCADA hardware and software components. It is responsible for controlling SCADA operational information, for example, controlling, observing and communication between several RTU and MTU in the form of text, statistics or other comprehensible content.\\
\textbf{Historian} is used for accumulating two-way communication data, events, and alarms between SCADA control centre. It can be described as a centralised database or a server located at a distant location. Historian is queried to populate graphical trends on the HMI.\\
\textbf{Communication network} provides communication services between various components in the SCADA network framework.  The medium utilised can be either wireless or wired. Presently, wireless media is generally utilised as it interfaces geologically circulated areas and less available zones to communicate effortlessly \cite{Pathak2014}. 
The advancement of communication paradigm is isolated into four primary ages, for example, the First era: Monolithic, Second era: Distributed, Third era: Networked, Fourth era: Internet of things technology.

\begin{enumerate} 
    \item \textbf{Monolithic SCADA systems: } \textcolor{black}{It refers to those systems which work in an isolated environment and do not have any connectivity to the other systems.} The motive of these systems is to work in a solitary way. Large minicomputers were used for SCADA system computing. PDP-11 series which was developed by Digital Equipment Corporation is an example of a first-generation SCADA system.
    In this architecture,  RTUs communicate to MTU using Wide Area Networks (WAN) as shown in Fig. \ref{fig:monolithicSCADA}. However, the WAN protocols used that time were in the preliminary stage.    
    
    The communication protocols were proprietary which can be used only with proprietary MTU from the same vendor. These protocols were limited to permit scanning, control and data exchange between MTU and RTUs. The interconnection between MTU and RTUs was done at the bus level \cite{SCADAGeneration}. Connecting different vendor RTUs to MTU was an impossible task resulting in an urgent requirement for the open standard. In some case, to provide redundancy to the SCADA system, an equally equipped system, working as a backup system was connected to the master system.

    \item    \textbf{Distributed SCADA systems: } These systems were inter-connected and confined inside small range network like Local Area Networks (LAN) as shown in Fig. \ref{fig:distributedSCADA}. This generation distributes the computation overhead on remotely located systems using LAN, i.e., some of the systems work as communications processors, some as operator interfaces, some as a database server, etc. \cite{SCADAGeneration}, resulting in more processing power, redundant, and reliable system. The Distributed architecture is used in case of multiple clients and stations. Similar to the monolithic SCADA, distributed SCADA systems were also confined to proprietary hardware, software, network protocols and peripheral devices that were supplied by the vendor \cite{NationalCommunicationsSystem2004,Meghanathan}. Security of SCADA systems was not of concern. The information was shared using the LAN. However, some of the LAN protocols used were proprietary nature which again kept a restriction on the systems which can be connected to a LAN to work as a distributed MTU. WAN was used to intercommunication between RTUs and MTU.   
    
    \item \textbf{Networked SCADA systems: } It utilises networks and web broadly because of the standardisation and cost-effective solutions for large-scale systems. This is also referred to as a modern SCADA system \cite{Pathak2014}. In this design, SCADA systems may be geographically distributed. However, Networked SCADA is closely related to that of Distributed SCADA, with the significant difference in the usage of open protocols and standards for communication rather than proprietary protocols resulting in distributing MTU functionality across a WAN also as shown in Fig. \ref{fig:networkedSCADA}. Due to the usage of open standards, third-party peripheral devices can be connected to the network \cite{SCADAGeneration}. The significant game-changing improvement in networked SCADA was the use of Intenet Protocol for the communication between MTU and RTUs, resulting in disaster survivability.
    
     \item \textbf{Fourth generation:} The industries have been utilising the power of technology to build, monitor and control the systems. Integration of Internet of Things (IoT) innovation and economically accessible cloud computing with SCADA systems has considerably lessened its infrastructure and deployment costs. Moreover, the integration and maintenance are also easy as compared to the previous generations \cite{forthgen}. 
     Industries 4.0 is an example of a fourth generation SCADA system as shown in Fig. \ref{fig:fourth_gen_SCADA}. It includes distributed cognitive computing, CPS, IoT, and cloud computing \cite{Industry4}. SCADA systems already share a few characteristics of IoT, e.g. data access, manipulation and visualisation. IoT differs in terms of interoperability, scalability and capability of big data analytics. The collection and control of all data are done using an open communication standard. The collected data is stored on clouds and extraction to get valuable insights from data. Industrial Internet of Things (IIoT) or Industry 4.0 refers to the developments in fourth generation SCADA systems. IIoT is described as IoT in industries. It is a network of devices with a significant focus on transfer, control of critical information, getting insights from large data,. Therefore, to inculcate IIoT in SCADA, several devices, protocols need to be integrated into the existing system. \textcolor{black}{IIoT has also improved its resilience by identifying anomalous behaviour using data-driven techniques \cite{Yang2014,Feng2019, Lyu}. Also, the CI system has a significant concern for the losses due to downtime. However, using predictive maintenance, these downtimes can be reduced, increasing the production of the system \cite{Zhang2014DataMA}.}       % needed in the second column of the first page if using \IEEEpubid
    %\IEEEpubidadjcol
    \begin{figure*}[t!]
	\centering
	\includegraphics[width = 1\textwidth, height =9cm]{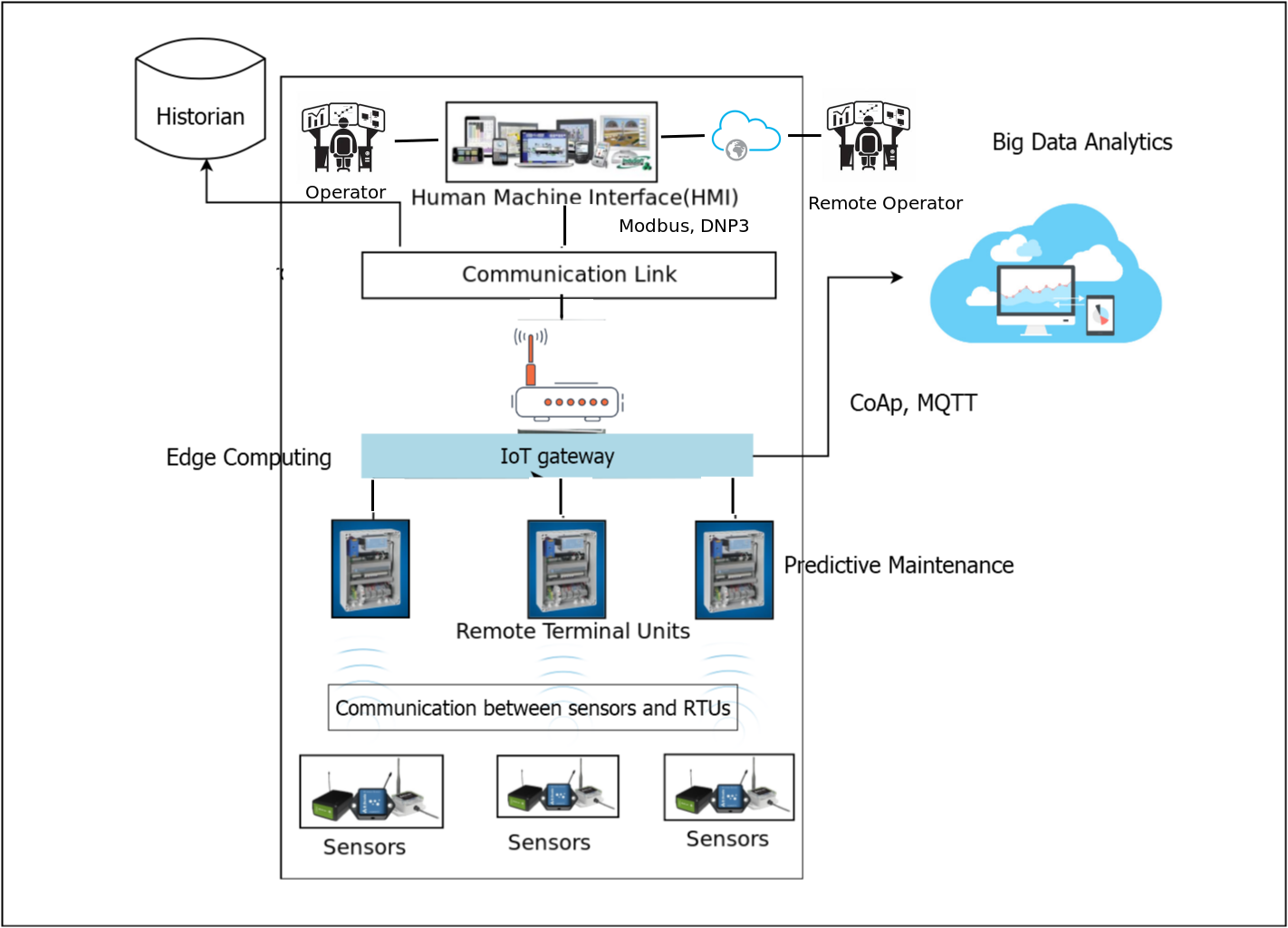}
	\caption[]{\textcolor{black}{Fourth generation SCADA}}
	\label{fig:fourth_gen_SCADA}
\end{figure*}

     \begin{table*}[h!]
        \centering
        \caption{\textcolor{black}{Comparison of various communication protocols}}
        \label{tab:Comparision_protocols}
        \begin{tabular}{p{1.5cm}p{2cm}p{2cm}p{2cm}p{2cm}p{2cm}p{2cm}}
            \noalign{\hrule height 2pt}
            \bf    Attribute & \bf Modbus  &\bf DNP3 & \bf IEC 6870-5-101 & \bf Foundation Fieldbus & \bf Profibus  & \bf IEC 61850 
            \\\noalign{\hrule height 2pt}
            Year & 1979 & 1993  &1995 & 2004 & 1989& 2005 (Project started in 1995)
            \\\hline
            Organization &    Gould Modicon &     Harris, Distributed Automation Products &    IEC & FieldComm Group  &    Promoted by BMBF (Germany) &  IEC Technical  Committee  57
            
            \\\hline
            Number of layers &    1 &    4       & 3 & 4  & 3 &3
            \\\hline
            Architecture &    Single layer i.e. Application layer    & 4 layer architecture & 3 layer architecture based on EPA model.
            &     4 layer architecture  &    3 layer architecture &  3 layer architecture
            \\\hline
            Addressing &    8-bit address &    16-bit source and destination addresses &    0, 8, 16-bit addresses are supported & 8, 16, 32-bit addresses are supported &    7-bit address
            (0-3 address are used by master and rest by slaves)  &  48-bit source and destination addresses
            \\\hline
            Users &    Target low volume data applications &    China, North America, and  Australia  &    Europe, China & America and France &    All over the world & All over the world
            \\\hline
            Source    & Open source &    Open source    &    Commercially available & Open source  & Commercially available & Open source
             \\\hline
             \textcolor{black}{
            Security state  }  & \textcolor{black}{No encryption and authentication control}  &  \textcolor{black}{DNP3-SA support encryption and authentication control }  & \textcolor{black}{ No encryption but supports authentication control}  & \textcolor{black}{No encryption and authentication control} & \textcolor{black}{ Supports encryption and authentication control} & \textcolor{black}{No encrption but supports authentication control}
            \\\noalign{\hrule height 2pt}
            
        \end{tabular}
        
    \end{table*}
    \textcolor{black}{ }
\end{enumerate} 
% A detailed survey of SCADA communication protocols is done in next subsection.
\subsection{SCADA Communication Protocols}
The communication protocols are regulations for the data depiction and exchange over a communication link. \cite{Perera}. SCADA communication protocols play a pivotal role in MTU-RTU interactions. At first, instruments and protective relays permitted remote communications using local RS232 association or using a dial-up modem interface. But due to scalability issues, they have moved to more advanced protocols  \cite{uzair}. 
\par  As the SCADA system is a composition of many components, if each component uses a vendor-specific protocol, it will not be able to communicate with other components. 
Each vendor-specific SCADA protocol has its own rules and procedures of communication which can vary from data presentation and conversion, assignment of addresses to command generation and status information. Therefore, to support interoperability and cost efficiency, some open standards were presented.
\par
To encourage open protocols, the Open System Interconnection (OSI) Model was introduced in 1984 \cite{Companies2006}. The OSI model shows the data communications process composed of seven independent layers, and each of the layers describes how the data is handled in the different stages of transmission. Open protocols increase the availability of the device, interoperability, vendors independence, optimised cost, easy technical support, etc. \\
A study of various communication protocols is done below.
\begin{enumerate}
    \item \textbf{ Modbus:} The Modbus transmission protocol, an application layer messaging protocol was developed by Gould Modicon for their Modicon programmable controller \cite{WINGPATH}. It is the most commonly used protocol for connecting the electronic devices due to openly published and easy to use. Moreover, it is used for the interactions between MTU and RTUs.
    \par
    A typical Modbus network supports one master and a maximum of two hundred forty-seven slaves.  RTUs only reply to messages targeted to them but avoid responding to the broadcasts \cite{RDMS}.
    It uses four types of communication messages such as request/response message to/from MTU, acknowledgement message for the successful delivery of the message at the MTU and RTUs. MTU can send messages to the slaves and also assign an address to each of the slaves which vary from 1 to 247. Modbus/TCP, an enhanced variation of Modbus is also available which focus on reliable communication over the Internet and Intranet. It follows TCP/IP's error detection methods to detect the errors.\par
    Modbus plus protocol is proposed to overcome the master terminal vulnerability issues. It is a token-based protocol. Modbus protocol assembles the request message transmitted from the remote terminal to the master terminal into PDU which is an amalgam of the data request and a function code. PDU changes over into an application information unit by including function code fields at OSI layer. Similarly master terminal will send a reply to the remote terminal. However, due to extra cable and other communication issues, it is not preferred for real-time communication. \par
    
    \item\textbf{ DNP3:} Distributed Network Protocol (DNP) protocol is based on the Enhanced Performance Architecture (EPA) model. EPA is a streamlined type of OSI layer architecture. It was developed by Harris, Distributed Automation Products  \cite{Shahzad2014a}.  The motive for DNP3 protocol development was to obtain open and standards-based interoperability between RTUs, MTU, and Programmable Logic Controller(PLC).
     \par Data link layer convention, transport functions, application conventions and data link library are the core components of the DNP3 protocol. A user layer is appended to the EPA architecture which is responsible for multiplexing, data fragmentation, prioritisation and error checking, etc.    In the layered architecture of DNP3 protocol, application layer details the packet design, services, and procedure for the application layer. This message is then forwarded to the pseudo-transport layer which forwards the segmented data unit to the data link layer \cite{uzair}. It further forwards the message to the physical layer \cite{Mahapatra2015}. It supports multiple-slave, peer-to-peer (P2P) and multiple-master communication.

    \item \textbf{ IEC 60870-5 Protocol:} The International Electro-Technical Commission(IEC) 60870-5 protocol also follows EPA model. 
    The application layer is included as an additional top layer of EPA architecture which indicate the functions related to telecontrol framework. Telecontrol framework based variations e.g. T101, T102, T103, T104 characterize diverse particulars, data objects and function codes at application convention level \cite{IPCOMM}.
    For the efficient transmission, DNP3 layer stack adds a pseudo-transport layer, but it is not used in IEC 60870-5.
    
    \item\textbf{Foundation Fieldbus Protocol:} This protocol was presented by FieldComm Group \cite{Wikipedia}. The user, application, data link and physical, the four-layer stack is used in Foundation Fieldbus. The architecture of Foundation Fieldbus follows the OSI layer model in which the user layer is added as an additional top layer of the application layer. The user layer acts as a gateway between software programs and field devices. Easy process integration, multifunctional devices, open standard, decrease massive wire cost features superior it from other protocols.

    \item \textbf{ Profibus Protocol:} Process Field Bus (Profibus) protocol was promoted by BMBF (Germany). The communication of data between MTU and RTUs is a cyclic process. MTU reads RTUs input data and writes RTUs output data. Field Bus Message Specification (FMS), Distributed Peripheral (DP), and Profibus Variations (PA) are the three versions of Process Field Bus (Profibus) protocol. Profibus is most popularly used in discrete manufacturing and process control \cite{Shahzad2014a}.
      \item \textbf{ IEC 61850 Protocol:}  The International Electro-Technical Commission(IEC) 61850 protocol was developed by the IEC Technical  Committee  57 \cite{IEC61850_wiki}. A group of manufacturers  (ABB, Alstom, Schneider, SEL, Siemens, Toshiba, etc.) proposed this protocol to improve the interoperability of equipment \cite{Czechowski2015CyberSI}.  This protocol differs from other OSI reference models in the sense that it also describes how data is executed and stored apart from how it is sent and received. The source and destination address are 48 bits each \cite{IEC61850}. IEC 61850 is generally used in electrical substations for communication among intelligent electronic devices \cite{IEC61850_wiki}. Moreover, IEC 61850 abstract data models can be mapped to many other protocols, e.g. MMS, GOOSE, and SMV \cite{Yang2017}.  
%    IEC 61850 abstract data models can be easily mapped to other protocols.
\end{enumerate}

Accordingly, SCADA communication conventions have advanced from restrictive to business/open source conventions. SCADA framework's unwavering quality relies on its correspondence conventions. A brief and comparative analysis of communication protocols available for SCADA is Table \ref{tab:Comparision_protocols}. Since DNP3, IEC 60870-5-101 and Foundation Fieldbus are open Standards \cite{Choi2013}. These protocols are more widely used. DNP3 and IEC 60870-5-101 focus on providing the first level solutions of Data Acquisition Interoperability.  These are required to communicate outside the substation \cite{Mahapatra2015}.  DNP3 allows SCADA systems to poll at different frequency while IEC 60870-5-101 poll at the same frequency which helps it is a case of limited bandwidth. The packet size in DNP is larger than IEC 60870-5-101. Hence for long distance DNP3 protocol is favoured.  
Modbus is, for the most part, utilised for applications where the volume of information exchange is low \cite{uzair}. It is a quick and safe convention, and a ton of data is loaded in one message\cite{Perera}. Modbus is a single layer protocol while DNP3, Foundation Fieldbus uses four-layer architecture. Modbus is mainly targeted for low volume data applications. \textcolor{black}{ Only DNP3-SA and Profibus support encryption and authentication control, while Modbus is an insecure communication protocol. IEC-6870-5-101  and IEC 61850 do not support encryption but allow authentication control. }
Many factors affect the protocols selection for communication, for example, the utility of the system, location where the SCADA system will be implemented.  Since choosing the best protocols ensures that if needed the developed system will have good potential for scalability. Systems should have the flexibility to incorporate security in communication protocols. 
\par Apart from these traditional communication protocols, in IIoT based SCADA other IoT protocols, e.g. Zigbee, Bluetooth Low Energy (BLE), Long Range (LoRA) etc. are used for communication. 
\begin{enumerate}
    \item \textbf{Zigbee:} Zigbee, an IEEE 802.15.4 based communication protocol, is developed by Zigbee alliance. Zigbee was standardised in 2003, and revised in 2006. The range of Zigbee communication is between 10 to 100 meters line-of-sight depending on environmental characteristics. Zigbee architecture includes three types of devices, i.e. Fully Functional Device (FFD) (act as a router), Reduced Functional Device (RFD), and a coordinator. It enables  Wireless Personal Area Networks (WPAN) and provides a communication protocol with low power digital radios. 
    In short, it is a low data rate, low-power and low communication range wireless ad hoc network which is secured by 128-bit symmetric encryption keys and data rate of 250 kbps. 
    \item \textbf{Bluetooth :} Bluetooth special interest group developed with a motive to decrease the power consumption as compared to classic Bluetooth technology. The protocol stack in BLE is the same as in classic Bluetooth. BLE supports a quick transfer of small data packets with 1 Mbps data rate. It does not support data streaming. It follows master-slave architecture. Masterbehave like a central device which connect to many slaves, resulting need of these devices power-efficient. The energy is saved by keeping the slave nodes in sleep mode by default and wake up these nodes periodically to send data packets to the master node and receive control packets from the slave node.  BLE is 2.5 times energy efficient than Zigbee \cite{Sethi2017InternetOT}. 
    \item \textbf{LoRA :} LoRA, a long range communication protocol, was developed by Cycleo of Grenoble, France. In 2012, it was acquired by Semtech. It supports long-range communication up to 10 Km and data rate less than 50kbps with low power consumption. It is most suitable for non-real time application which is fault tolerant. It works in the physical layer combined with  Long Range Wide Area Network, in the upper layers. 
\end{enumerate}
 \begin{table*}[b!]
    \centering
    \caption{\textcolor{black}{Threats to the SCADA systems}}
    \label{tab:threats_SCADA}
    \begin{tabular}{ p{3.65cm} p{8.5cm} p{3cm}}
        \noalign{\hrule height 2pt}
        \bf    Threats & \bf Description  & \textcolor{black}{\bf Vulnerable SCADA Component}
        \\\noalign{\hrule height 2pt}
        Physical security  & SCADA systems are geographically distributed. Hence their physical security is a big issue \cite{HyungJun, Markovic}. & \textcolor{black}{ All components}
        \\\hline
        Operating System Vulnerabilities &    The SCADA system is expected to be running continuously without interruption. So any patch to the SCADA system cannot be applied. & \textcolor{black}{All components}
        \\\hline
        Authentication Vulnerabilities, i.e., Permission, Privileges, and Access controls &    Generally, for employee convenience, the passwords are shared, which eliminates the sense of authentication and accountability \cite{threatsSCADA}. Also, some vendors put default passwords, which are used without modification by the user. Moreover, password policies also very weak \cite{Acquisitiona}.   & \textcolor{black}{ All components }
        \\\hline
        Improper authentication, i.e., Unauthorized remote access  &    Due to the geographic distribution, to monitor the system, remote access is required. Remote access is more vulnerable to unauthorized access. & \textcolor{black}{All components}
        \\\hline
        % Complex interconnections  &    Now, the SCADA systems are much complex consisting of many interconnections. The weakness of the framework is specifically relative to the number of open ports. & HMI, RTU, PLC and MTU
        % \\\hline
        Audit and Accountability, i.e., Monitoring and Defenses &    Cryptographic communications, Intrusion detection system (IDS), firewall are not universally used. It is challenging to implement these cryptographic approaches on sensors or actuators, considering the resource capability and scale \cite{threatsSCADA}.  Security documentation is also limited. The potential for zero-day attacks is always present. The security assessments tools are also lacking to achieve up to the mark performance. & \textcolor{black}{All components}
        \\\hline
        Wireless communication network    & In SCADA systems, the communication link is mainly wireless.  Depending on the implementation these links are vulnerable to the security attacks. & \textcolor{black}{All components }
        \\\hline
        Legacy SCADA Software    & Most of the SCADA systems use legacy software which was designed long ago.  Security of the system was not a consideration at that time \cite{threatsSCADA, Acquisitiona}. & \textcolor{black}{ All components}
        \\\hline
        Upgrade restriction    & The processors are constrained by low computation power and memory resources, and also these systems are not compatible with upgrades \cite{Acquisitiona}.& \textcolor{black}{All components}
        \\\hline
        Public Information    & In most of the application sectors, the design and architecture of SCADA system are published making it available to attackers. Also, employees working on a firm leak the information from their past working place \cite{Markovic}. & \textcolor{black}{All components}
        \\\noalign{\hrule height 2pt}
    \end{tabular}
    
\end{table*}

Apart from these device-to-device communication protocol, other application layer protocol e.g. MQTT, Constrained Application Protocol (CoAP) and Message Queue Telemetry Transport (MQTT) are developed for IoT environment as HTTP, HTTPs are not suitable due to resource constraints. 
\begin{enumerate}
    \item  \textbf{CoAP : } CoAP,  a specialized Internet Application Protocol,  is an replacement of HTTP for resource constraint IoT based devices \cite{CoAP1, CoAp2}. Low overhead, multicast and ease to use are the basic pillar for IoT devices. IoT devices have much less memory and power supply in comparison to traditional Internet devices.  It uses an Efficient XML Interchanges (EXI) data format that leads to a more space efficient protocol. It also supports resource discovery, message exchange, auto-configuration, built in header compression etc. It uses four types of message, i.e. confirmable, non-confirmable, acknowledgement and reset. Confirmation messages are used for reliable communication; acknowledge message is used for successful delivery of the message. By default, CoAP is bound to User Datagram Protocol (UDP) and security is provided using Datagram Transport Layer Security.  
    
     \item  \textbf{MQTT : } MQTT, a  publish-subscribe-based messaging protocol, was developed by IBM. It is client/server protocol, where clients act as a publisher or subscriber and server behaves like a broker.  The information is arranged in a topics hierarchy. Topics name are generally in text format, which increases the overhead. A client sends a control message to the server when it wants to publish a message. The server distributes the message to the subscribers later. Neither publisher nor subscribers need to share their configurations, location. MQTT is supported over Transmission Control Protocol (TCP), which restrict its use for all types of  IoT devices. MQTT control message size varies between  2 bytes to 256 megabytes. It supports 14 control messages to manage publisher-broker-subscriber communication \cite{MQTT}.
     \par Apart from MQTT, few extensions, e.g. MQTT-S/ MQTT-SN are proposed which specifically focus on cost and power effective solutions.   These include replacing topic text with topic Ids,  buffering procedure for nodes in sleep mode etc. MQTT-SN is proposed to use over UDP or Bluetooth.
\end{enumerate}

The communication network protocols do not support security features. Therefore, they are prone to cyber-attacks. In the next section, we discuss the inherent vulnerability of SCADA systems by looking at reported attacks. 
\section{Taxonomy of attacks} \label{Taxonomy of attacks}
Recently, the number of security-related attacks on SCADA system has drastically increased. Threats like Stuxnet \cite{Falliere2011}, Aurora \cite{McAfee2010}, Maroochy \cite{Slay2007} give us a clear idea of how much damage a determined adversary can cause even on the general public.

\par 
Table \ref{tab:threats_SCADA} summarises the various threats to the SCADA systems. \textcolor{black}{ Table 2 summarises the various threats to the SCADA systems. The physical security of these systems remains a significant issue due to geographical distribution. These systems are expected to run without any interruption, so any patch or upgrade cannot be applied without compromising its productivity [32, 33]. Moreover, most of the communication happens on the wireless network, which makes it vulnerable to network security attacks [35]. The architecture and design of SCADA systems are available in the form of patents or publications,  which  make  it  accessible  to  hackers  [34]. We have also highlighted the vulnerable SCADA component w.r.t. each threat. Sensors and actuators are prone to physical security as they are generally deployed in remote areas. PLC, MTU, and RTUs still uses legacy SCADA software, and are restricted to update. Therefore, these are even prone to well-known vulnerabilities exploitations. }
A lot of attacks have been detected even with advanced security solution enforced in the system. The first known cyber-security attack occurrence including SCADA framework was in 1982, in which enemy implanted a Trojan in the SCADA framework that was responsible for controlling the Siberian Pipeline. A brief analysis of the reported attacks is given in the next subsection.

\begin{table*}[h!]
        \scriptsize
        \centering
        \caption{Some of the Important Attacks during 1982-2016}
        \label{tab:attacks_state}
        \begin{tabular}{p{3.5cm} p{2cm} p{3cm} p{3cm} p{3cm}}
            \noalign{\hrule height 2pt}
            \bf Attack title (Year) & \bf Country (Industry) & \bf Target   & \bf Impact & \bf Type
            \\\noalign{\hrule height 2pt}
           \bf Siberian Gas Pipeline Explosion (1982) \cite{Risidata} & \textcolor{black}{Russia (Petroleum)} & \textcolor{black}{Pipeline}   & \textcolor{black}{Financial Loss, System Damage} & \textcolor{black}{Malware}
    \\\hline
    \bf 
    Sellafield Nuclear plant system error (1991)\cite{Risidata} &
   \textcolor{black}{ Uinited Kingdom }
    \textcolor{black}{(Power and Utilities) }&
    \textcolor{black}{Shielding Door } &
    \textcolor{black}{Production Loss} &
    \textcolor{black}{Noncyber attack}
    \\\hline
    \bf
    Virus in Nuclear Power Plant (1992) \cite{Risidata} &
   \textcolor{black}{ Lithuania(Power and Utilities)} &
    \textcolor{black}{Reactor}&
    \textcolor{black}{System Damage} &
    \textcolor{black}{Malware}
    \\\hline
    \bf 
    Hacking of Salt river project(1994 ) \cite{Risidata}&
    \textcolor{black}{United State}       
    \textcolor{black}{(Power and Utilities)} &
    \textcolor{black}{Software system} &
    \textcolor{black}{Financial Loss, DL}
    &
   \textcolor{black}{ Unauthorised Remote Access}
    \\\hline
    \bf 
    Worcester Air Traffic System Hack (1997) \cite{Risidata}&
    \textcolor{black}{United State} 
    \textcolor{black}{(Transportation)} &
    \textcolor{black}{Control System} &
    \textcolor{black}{System Damage}
    &
    \textcolor{black}{Noncyber attack}
    \\\hline
    \bf
    Maroochy (2000) \cite{Slay2007}&
    \textcolor{black}{Australia (Sewage Control System )}&
    \textcolor{black}{Flood gate}&
    \textcolor{black}{Environmental Damage}
    & \textcolor{black}{Unauthorised Remote Access}
    \\\hline
    \bf 
    Utility SCADA system attack (2001) \cite{Risidata}&
    \textcolor{black}{United State  (Power and Utils)}&
    \textcolor{black}{SCADA control system} &
    \textcolor{black}{System Damage, Financial Loss}
    &
    \textcolor{black}{Unauthorised Remote Access}        
    \\\hline
    \bf 
    SQL Slammer (2003) \cite{RosslinJohnRobles2008}&
    \textcolor{black}{United State (Petroleum)} &
    \textcolor{black}{Automation Segment} &
    \textcolor{black}{Daniel-of-Service}
    &
    \textcolor{black}{Interruption of Service}    
    \\\hline
    \bf 
    Virus injected in CSX train signaling system (2003)     \cite{Miller:2012:SSC:2380790.2380805} &
    \textcolor{black}{United State (Transportation)} &
    \textcolor{black}{Signal dispatching system} &
    \textcolor{black}{Latency} &
    \textcolor{black}{Malware}
    \\\hline
    \bf 
    Nuclear plant slammer attack (2003) \cite{Risidata}&
    \textcolor{black}{United State (Power and Utilities)} & \textcolor{black}{Nuclear power plant} &
    \textcolor{black}{System Damage, Financial Loss} &
    \textcolor{black}{Malware}
    \\\hline
    \bf  
    Nachi worm on control servers (2003) \cite{Risidata}&
    \textcolor{black}{France (Chemical)} &
    \textcolor{black}{Advanced process controller (APC)} &
    \textcolor{black}{Latency} &
    \textcolor{black}{Malware}    
    \\\hline
    \bf 
    Sessor worm (2004) \cite{Risidata}&
    \textcolor{black}{United State (Chemical)} &
    \textcolor{black}{Decentralised control system (DCS)} &
    \textcolor{black}{System Damage} &
    \textcolor{black}{Malware}
    \\\hline
    \bf 
    Sessor worm (2004) \cite{Risidata}&
    \textcolor{black}{United Kingdom (Transportation)} &
    \textcolor{black}{Check-in controller system} &
    \textcolor{black}{Latency} &
    \textcolor{black}{Interruption of Service}
    
    \\\hline
    \bf 
    Water company hack in Pennsylvania (2006) \cite{Risidata}&
    \textcolor{black}{United State (Water/Waste Water)} &
    \textcolor{black}{Water plant computer system} &
    \textcolor{black}{System Damage} &
    \textcolor{black}{Unauthorised Remote Access}
    
    \\\hline
    \bf  
    Phishing attack (2007) \cite{Risidata}&
    \textcolor{black}{Unk (Power and Utilities)} &
    \textcolor{black}{Employee computer} &
    \textcolor{black}{System Damage} &
    \textcolor{black}{Malware}
    \\\hline
    \bf 
    Emergency siren Activation (2008) \cite{Risidata}&
    \textcolor{black}{United State (Other)} &
    \textcolor{black}{Emergency Siren} &
    \textcolor{black}{Daniel-of-Service, System Damage} &
    \textcolor{black}{Interruption of Service}
    
    \\\hline
    \bf  
    Road Sign Hack (2009) \cite{Risidata}&
    \textcolor{black}{United State (Transportation)} &
    \textcolor{black}{Digital Road Sign} &
    \textcolor{black}{None} &
    \textcolor{black}{Unauthorised Remote Access}
    \\\hline
    
    \bf 
    Power Company Hack in Texas (2009) \cite{Risidata}&
    \textcolor{black}{United State (Power and Utilities)} &
    \textcolor{black}{Energy forecast system} &
    \textcolor{black}{Financial Loss} &
    \textcolor{black}{Unauthorised Remote Access}
    \\\hline
    
    \bf 
    Stuxnet (2010) 
    \cite{Yampolskiy:2013:TDC:2461446.2461465,Chen:2011:LS:1978246.1978332,Falliere2011,Langner:2011:SDC:1990763.1990881}&
    \textcolor{black}{Iran (Power/Utilities)}&
    \textcolor{black}{Centrifuges PLCs} &
    \textcolor{black}{System Damage, Financial Loss} 
    &
    \textcolor{black}{Unauthorised Remote Access}
    \\\hline
    \bf 
    South Houston Water Treatment Plant Hack (2011) \cite{Risidata}&
   \textcolor{black}{ United State (Water/Waste Water)} &
    \textcolor{black}{Plant controller} &
   \textcolor{black}{ None} &
   \textcolor{black}{Unknown}
    \\\hline
    \bf Auto manufacturer hacked (2012) \cite{Risidata}&
    \textcolor{black}{United State (Automotive)} &
    \textcolor{black}{Company computer} &
    \textcolor{black}{Intellectual loss} &
    \textcolor{black}{Malware}
    
    \\\hline
    \bf 
    New Year Dam attack (2013) \cite{Risidata}&
    \textcolor{black}{United State (Water/Waste Water)} &
    \textcolor{black}{Computerized control of Dam} &
    \textcolor{black}{Intellectual loss, System Damage} &
    \textcolor{black}{Unauthorised Remote Access}
    \\\hline
    \bf 
    Godzilla Attack (2014) \cite{Miller:2012:SSC:2380790.2380805}&
    \textcolor{black}{United State (Transportation)} &
    \textcolor{black}{Sign Board} &
    \textcolor{black}{System Damage, Intellectual loss} &
    \textcolor{black}{Unauthorised Remote Access}
    \\\hline
    \bf 
    Steel Mill Cyber attack (2014) \cite{Lee2014}&
    \textcolor{black}{Germany (Metal)}&
    \textcolor{black}{Furnace} &
    \textcolor{black}{System Damage} &
    \textcolor{black}{Unauthorised Remote Access}
    \\\hline
    \bf  
    Ukrainian Power Outage (2015) \cite{Trisal,3ICSattack}&
    \textcolor{black}{Ukraine(Power and Utilities)} &
    \textcolor{black}{Computer network} &
    \textcolor{black}{System Damage} &
    \textcolor{black}{Malware}
    \\\hline
    \bf 
    Operation Ghoul (2016) \cite{Trisal,3ICSattack}&
    \textcolor{black}{Middle Eastern Countries(Cyber Security Company)}&
    \textcolor{black}{Computer system} &
    \textcolor{black}{Data Loss}&
   \textcolor{black}{ Malware}    
    \\
    \noalign{\hrule height 2pt}
    
        \end{tabular}
        
    \end{table*}

\subsection{Analysis of Attacks}  In 2017, the Repository of Industrial Security Incidents (RISI) database \cite{Risidata}, a publicly available online database, contains 242 incidents that are recorded from 1982 to 2017. This data set is considered as one of the richest to date to understand the attacks taxonomy. \textcolor{black}{The real count of such attacks is much more than because many real-time attacks are not reported \cite{Miller:2012:SSC:2380790.2380805}.} \par
\begin{figure*}[h!]
    \centering
    \begin{minipage}{.5\textwidth}
        \centering
        \includegraphics[width=2.75in, height =3in]{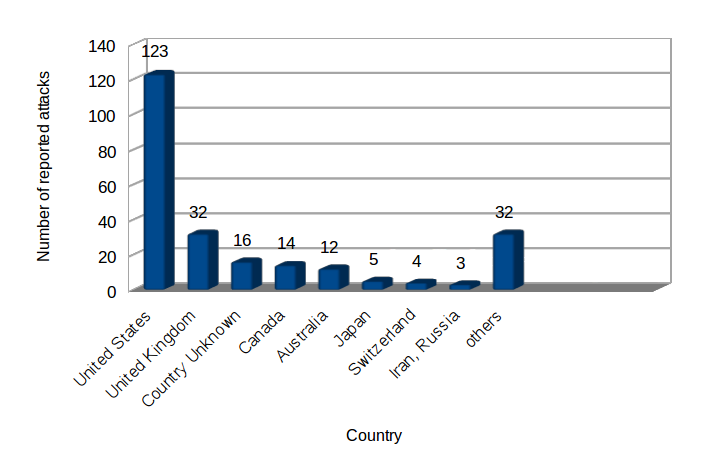}
        \caption{Statistical view for Country vs Attacks count }
        \label{fig:Statistical_view_Threats_SCADA_country}
    \end{minipage}%
    \begin{minipage}{.5\textwidth}
        \centering
        \includegraphics[width=2.75in, height =3in]{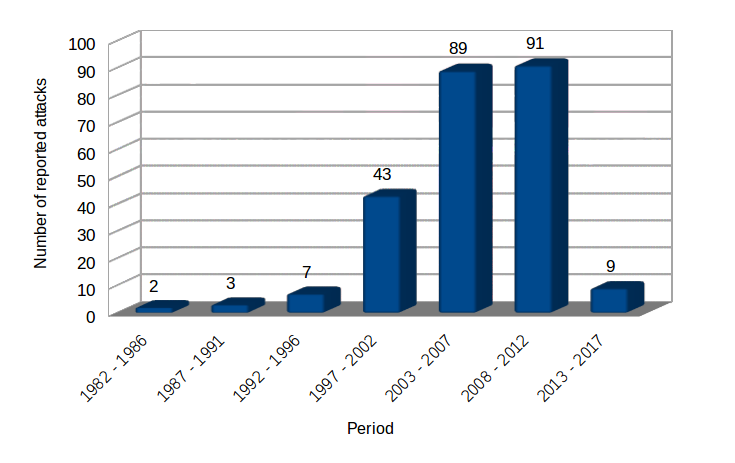}
        \caption{ Statistic view for Period vs Attacks count }
        \label{fig:Statistical_view_Threats_SCADA_period}
    \end{minipage}
    \begin{minipage}{.5\textwidth}
        \centering
        \includegraphics[width = 2.75in, height =3in]{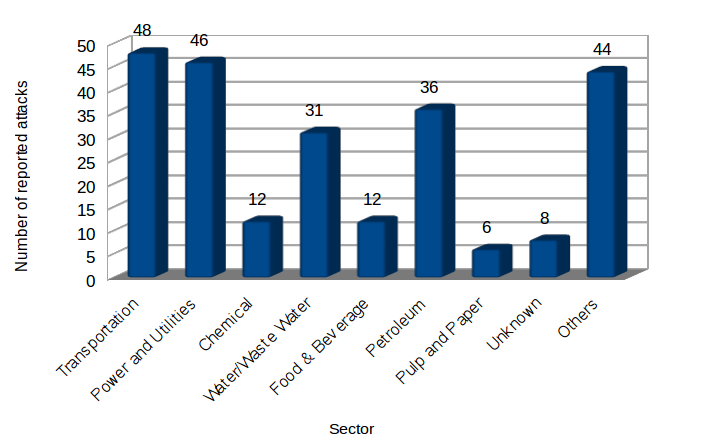}
        \caption{Statistic view for Sector vs Attack count }
        \label{fig:Statistical_view_Threats_SCADA_application_sector}
    \end{minipage}%
    \begin{minipage}{.5\textwidth}
        \centering
        \includegraphics[width=2.75in, height = 3in]{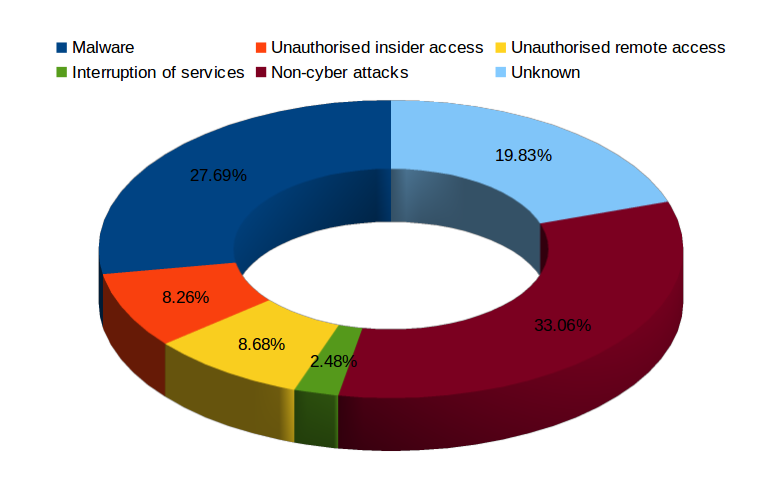}
        \caption{Threats statistic view for for Attack category }
        \label{fig:Statistical_view_Threats_SCADA_attack_category}
    \end{minipage}
\end{figure*}

 
It is necessary to analyse the previous security assaults to prevent future attacks, i.e., how the attacks have been carried out \cite{Miller:2012:SSC:2380790.2380805}? How can the system be made more robust against these attacks? 
Moreover, Henrie in \cite{doi:10.1080/10429247.2013.11431973} commented on the current cyber state of SCADA system that these attacks are "real and expanding". An in-depth analysis of these security incidents can provide the capability to detect and prevent these attacks priorly. Miller and Rowe analysed past attack records based on originating sector, the way attack was implemented, and attack target sectors. Their study on previous attacks gives the nature of those attacks. \\
 \textcolor{black}{ In Table \ref{tab:attacks_state}, We summarise some of the high-impact SCADA security incidents chronologically. The table highlighted the country and industry in which the attack was reported. It also lists the target component, impact of the attack, the method used to launch the attack.
The impact of the attack is categorised into six categories, i.e., Financial Loss, System Damage, Production Loss, Daniel of Service,  Latency, and  Intellectual loss. We further classified the type of attacks into five categories, i.e., Malware, Noncyber attack, Unauthorised Remote Access, Interruption of Service, and Unknown. Unknown denotes the attack category for which the source is still unknown. The attacks in Table \ref{tab:attacks_state} are chosen to cover a maximum number of impacted industries over the years. 
}
According to RISI repository \cite{Risidata}, about 17 countries have one reported security attack per country. 
The entire RISI dataset was analysed to find out patterns and highlight key points.  Organised hacking groups cause 5\% of the reported attack.  The result of the analysis in Fig. \ref{fig:Statistical_view_Threats_SCADA_country} number of reported attacks vs country shows that the USA and UK are the countries most affected by cyber-attacks. Sixteen reported attacks do not mention the country name. 

There are seven countries which have two cyber-attacks per country. However, this observation depends on the quality and completeness of the RISI database. The completeness of the RISI data set depends on the nations who report these attacks.  Moreover, 20\% of the attacks on critical infrastructure are unknown \cite{Ogie2017}. 
\par

In Fig. \ref{fig:Statistical_view_Threats_SCADA_period}  we analyse the number of attacks reported vs period. We have examined the count over a five-year interval. Most numbers of attacks, i.e., 91 have been published during 2008 -2012 followed by 2003- 2007. The highest number of attacks are reported in 2003 (36), most of which were due to malware attack.  However, after that, there is a decrease in the count of attacks. 
\par 
In Fig. \ref{fig:Statistical_view_Threats_SCADA_application_sector} we analyse which application sector is more prone to the attacks. Forty-eight attacks have been reported in the transportation sector which is followed by 46 attacks in power and utilities. The reason for the more vulnerable industry may depend on the revenue obtained due to the attack. Moreover, an attack can originate from many sources to harden the mitigation processes.

\par
Fig. \ref{fig:Statistical_view_Threats_SCADA_attack_category} shows that approximately 28\% of the reported attacks are due to malware attack. Unauthorised access is also another cause of many attacks. Therefore, adequate security policies should be practised in industries. 
Miller and Row in \cite{Miller:2012:SSC:2380790.2380805}, analysed that there is a drastic increase in the count of cyber-attacks over the years. As per the Dell annual report \cite{Dellreport}, the number of attacks against SCADA systems doubled in 2014 on the year-to-year basis.  The expert also confirmed that most of these attacks are politically motivated. The countries which have extensive SCADA systems are  Finland, the United Kingdom, and the United States. We need to strengthen cyber-security measures of SCADA systems, to shield them from cyber assault \cite{LUIIJF2011124,CHEN20169}.
\begin{figure}[h!]
    \centering
    \begin{minipage}{0.5\textwidth}
        \centering
        \includegraphics[width=2.65in, height = 2.25in]{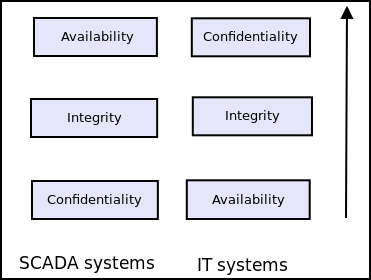}
        \caption{Priority order for General IT and SCADA}
        \label{fig:priority_SCADA}
    \end{minipage}
\end{figure}
The attacks on SCADA have miserable effects. New secure architectures are required for SCADA systems. Therefore, Fawzi et al. in \cite{Fawzi2014a} proposed a model embedded in control theory to guarantee appropriate "estimation and control of linear systems" when an aggressor degrades a portion of the sensing devices. The strength of this approach to accurately reconstruct the states during attacks is ably demonstrated utilising discrete simulations. Cardenas et al. first explored research challenges for the security of the cyber-physical system (CPS). \cite{Cardenas:2008:RCS:1496671.1496677}. The authors focus on the requirement of secure CPS and also discussed some of the vulnerability that might occur due to the fusion of cyber and physical systems.
Clifford Neuman, in \cite{Neuman2009}  focus on the design for the secure CPS. He has also enlightened the possible research area that will enhance the security of the CPS. In the proposed work, the author suggests combining security as an integral part of the basic design of CPS. For SCADA system, the security goal is generally reverse of the prioritised security goals for traditional information technology systems, as shown in Fig. \ref{fig:priority_SCADA}. Therefore, attackers generally target to interrupt the SCADA system availability.   \par  With time, attackers have started using more sophisticated techniques to compromise the security of SCADA systems than ever, so the threats are increasing. An attack scenario using electric vehicle infrastructure is described in \cite{6400439}. Till now, attackers have mainly focused on high-level systems, i.e., HMI and communication protocols. Surprisingly, the exploitation of field device firmware is a least focused research area \cite{SCHUETT201461,BASNIGHT201376,ZHU201726}.

Hence a quick and efficient attack detection systems are required, and we will discuss attack detection systems in the next section.

\section{Intrusion Detection Systems} \label{Intrusion Detection}
NIST \cite{nist} characterises IDS as the procedure of observing events in a host system or network, and these events are analysed for signs of unusual incidents. \cite{Singhal2010}. IDSs monitors the traffic and operation of network and host system; if it senses some security violation, the system administrator is notified. The research work for IDSs has been carried out since the 1980s by Aderson. Generally, for analysing system behaviour, IDSs need training and validation data sets of anomaly and attacks. The research work for the IDSs suffers from the lack of datasets for the verification of the functionality of their algorithms. \par We surveyed some of the widely used publicly available datasets. Power system dataset \cite{powersystemdataset} include measurements related to the electronic transmission system, control, cyber-attacks behaviour collected from Snort \cite{snort}.
Gas pipeline and water storage tank dataset \cite{gassystemdataset,gassystemdataset1} consist of cyber-attacks against two lab scale frameworks. This was created using re-enactment of actual defective and ordinary operations of a gas pipeline and water tank separately. It consists of three categorical features which include payload info, ground truth, and network info. 2,74,623 instances with twenty-row features have been involved in this dataset. Moreover, some unusual pattern were identified in this dataset which helps to machine learning algorithms to detect attacks.
KDD99 \cite{Tavallae} is widely used dataset since 1999 for the evaluation of IDSs. It is created by using data collected in DARPA'98 IDS. It consists of forty dimensional 49,00,000  single connection records. However, this dataset does not include analytical or experimental validation of data's false alarm characteristics.  It also includes redundant and duplicate instances. Therefore, a re-sampled version of KDD dataset, NSS-KDD \cite{zgr2016ARO} dataset was created.
The first DARPA dataset, simulated over an air force base, was published by MIT Lincoln Lab in 1998 \cite{DARPA}. However, in 1999, an improved version of this dataset which includes suggestion by computer security communities was released, This dataset provides raw host and network dataset which need to be preprocessed to use for verifying machine learning IDSs. Apart from the above-discussed databases, National Vulnerability Database (NVD) \cite{nvd_database}, an extensive and publicly available database for the software and hardware vulnerabilities in a different domain is a good source for extracting SCADA specific vulnerability. NVD includes an examined analysis of all these reported vulnerabilities using the Common Vulnerability Scoring System (CVSS) framework and provides a base severity score for vulnerability by considering the scope of the attack, vulnerability component, impacted component, attack vector and complexity, frequency, privilege required etc. NVD indexes reported the vulnerability to Common Vulnerability Enumeration (CVE) \cite{CVE} Ids that enable automated vulnerability management. CVE Ids help to provide a common name for publicly available vulnerabilities. 
However, a lack of the complete SCADA attack data sets inhibits cybersecurity research for SCADA. There is not a comprehensive dataset covering all the attacks worldwide. Therefore, researchers are required to create the datasets by simulating test-bed with attacks. Moreover, only a few algorithms exist for datasets creation. For zero-day attack detection, advancement in these algorithms is required \cite{Rodofile:2017:FSC:3014812.3014883}. Rodofil et al. in \cite{Rodofile:2017:FSC:3014812.3014883} proposed a modular dataset generation framework for SCADA cyber-attacks. Yang et al. in \cite{Yang2011}, simulated the influence of a simple cyber attack in smart grid compromising the integrity of the system. Authors highlighted an immediate need to look for a robust and timely technical solution for detecting and preventing cyber attacks.
\par 
An IDS consist of sensors, analysis and detection engine, a notification system. Sensors which are deployed either on host or network are responsible for collecting network and host data. The received data is sent to the analysis and detection engine, which investigate and detect the presence of intrusions. If an intrusion is detected, a notification system notifies the system administrator. 
IDS techniques can be studied based on the source of information, and analysis methodology. 
A brief analysis of these detection techniques is given below.

\subsection{Classification Based on the source of information}
Based on the source of information, IDSs are generally divided into Host-Based Intrusion Detection(HIDS)
% , Distributed Intrusion Detection System(DIDS),
and Network Intrusion Detection System(NIDS). 
HIDS relies on the host activity and states information which can be file-system modifications, application logs. To specify/detect host-level misbehaviour is easy as HIDS auditing is distributed \cite{Mitchell2014}.
NIDS relies on the traffic generated on the network by the various set of devices. 
% DIDS relies on the data coming from different IDSs that are spread over a vast network. DIDS allows instant attacks data, advanced network monitoring, and attack incident analysis. 
\par 
Table \ref{tab:Comparision_type_IDS} shows an analysis of classification of IDSs. HIDS provides approximate real-time intrusion detection without requiring extra equipment. 
\begin{table*}[h!]
    \centering
    \caption{\textcolor{black}{Comparison of various type of IDSs}}
    \label{tab:Comparision_type_IDS}
    \begin{tabular}{ p{2.8cm} p{2.8cm} p{4cm} p{3cm}p{1.8cm}}
        \noalign{\hrule height 2pt}
        \bf Intrusion detection systems (IDSs) classification &\bf Input  &\bf Advantages &  \textcolor{black}{\bf Limitations }& \bf Examples
        \\\noalign{\hrule height 2pt}
        \multirow{5}{10cm}{Host Intrusion \linebreak Detection System \linebreak(HIDS)}&     \multirow{5}{10cm}{Relies on the host \linebreak activity  
            and states \linebreak information.}& 1. Lower cost of entry & \textcolor{black}{1. Fail to detect internal}  &    Tripwire \cite{tripwire}
        \\ 
        &&2. No additional hardware required.& \textcolor{black}{attacks and DoS.}&OSSEC \cite{ossec}
        \\ 
        &&3. Detect attacks that NIDS miss.& \textcolor{black}{2.The host, where } &
        \\ 
        &&4. Near-real-time detection and response.& \textcolor{black}{HIDS resides on}&
        \\
        &&5. Monitors specific system activites. & \textcolor{black}{is susceptible to  disablement} &
        \\\hline
        \multirow{5}{10cm}{Network Intrusion  \linebreak Detection System \linebreak (NIDS)} & \multirow{5}{10cm}{ Relies on the traffic \linebreak generated on the \linebreak network by various set \linebreak of devices.} 
        & 1. Real-time detection and response.
        & \textcolor{black}{1. It  fail to analyze  }& Snort \cite{snort}   
        \\ 
        &&2. Detect attacks that HIDS miss.& \textcolor{black}{ encrypted information.}&Bro \cite{bro}
        \\ 
        &&3. Independent from operating system.& \textcolor{black}{2. Fail to block the} &
        \\ 
        &&4. Removal of evidence of NIDS is difficult.&\textcolor{black}{attacks.}&
        
        \\
%         \hline
        
%         \multirow{5}{10cm}{Distributed Intrusion \linebreak Detection System \linebreak(DIDS)} & \multirow{5}{10cm}{Relies on the data com-\linebreak ing from various IDSs \linebreak that are spread over a \linebreak large network.} & Protect the global information infrastructure. & \multirow{5}{10cm}{Distributed Signature \linebreak based Intrusion}
        
%         \\ \cline{3-3}
%         &&Used for coordinated 
%         attacks involving multiple attackers that require global scope for assessment.&
%         \\
        \noalign{\hrule height 2pt}
        
    \end{tabular}
    
\end{table*}
HIDS such as Tripwire \cite{tripwire} and OSSEC \cite{ossec} uses whitelists of the filesystem. It is performing file integrity scans which identify any abnormalities which can classify possible intrusions. Moreover, NIDS provide real-time detection, and it is hard to remove evidence of NIDS. NIDS such as Snort \cite{snort}  and Bro \cite{bro}  use rule sets that define a type of intrusion or unacceptable behaviours such as a port scan or a DoS attack attempt. \textcolor{black}{ Shekari et al. in \cite{Shekari} proposed a  radio frequency-based distributed intrusion detection system (RFDIDS) for SCADA systems. Even if the entire SCADA system is considered untrusted,  RFDIDS  remains reliable.  The monitoring of the power grid substation activities is done using radiofrequency emissions (particularly at low frequencies ). 
Flosbach et al. in \cite{Flosbach} proposed an extensible and scalable network-based IDS to secure control networks in the domain of power distribution. They mainly targeted to detect process-based attacks, e.g. manipulated control commands by assessing the local physical process and all control commands continuously. They have also successfully deployed their model at a Dutch power distribution substation.   
}

% we propose a deep-learning-based network intrusion detection system for SCADA networks to protect ICSs from both conventional and SCADA specific network-based attacks. Instead of relying on hand-crafted features for individual network packets or flows, our proposed approach employs a convolutional neural network (CNN)to characterize salient temporal patterns of SCADA traffic and identify time windows where network attacks are present. In addition, we design a re-training scheme to handle previously unseen network attack instances, enabling SCADA system operators to extend our neural network models with site-specific network attack traces. Our results using realistic SCADA traffic data sets show that the proposed deep-learning-based approach is well-suited for network intrusion detection in SCADA systems, achieving high detection accuracy and providing the capability to handle newly emerged threats.

A consolidated DNP3 parser and validation policy are used in Wireless Bro to apprehend and handle the data communicated by SCADA devices. HIDS sensors are avoided to use in the SCADA components due to resource constraints. In comparison to HIDS , NIDS are generally preferred in SCADA networks. HIDS sensors cannot be installed owing to resource constrained of SCADA components. 
 \subsection{ Classification based on analysis strategy  }
In the analysis strategy, signature detection and anomaly detection are the major intrusion detection techniques. Apart from this, specification-based approaches are discussed under analysis strategy. An analysis of these approaches is discussed below.
 \\ \\ \textbf{Signature based intrusion detection technique: }
    In signature detection techniques, network traffic is matched with an attack signature, i.e., misuse pattern of the intrusive detection stored in IDS. The behaviour of the system is compared based on the attribute of the network traces. If any host or network activity matches with stored signatures, an alert is triggered. This approach can achieve a good accuracy for intrusion detections which depends on the correctness of the misuse pattern. This technique is effective to detect known attacks, but it fails to detect new attacks due to the absence of signature of new or variants of known attacks. Oman et al. in \cite{Oman2008}, presented a signature-based SCADA test setup for the power-grid sector to detecting the adversaries.  However, in this proposal, the automatic gathering is done only for RTUs. Yang et al. in \cite{Yang_rule_based_IDS}, proposed a,  rule-based IDS for IEC 60870-5-104 protocol. The abnormal events categorisation is done based on "Non-IEC/104 Communication","Spontaneous Messages Storm," Remote    Control Commands or Remote    Adjustment Commands from Unauthorized Client,"Reset Process Command from Unauthorized Client" and "Potential Buffer Overflow." Authors represented their approach using protocol traffic case-study. Anomaly detection systems can work efficiently if the traffic is regulated and have predictable behavior\cite{GENGE201524}.
\\ \\ \textbf{Anomaly detection based intrusion detection technique: } An anomaly detection, the system compares current network traffic with standard behaviour profile and if something (extremely) unusual appears then alert is raised. In this system, known intrusions are not required. The distinctive patterns are learned over time with specific statistical profiling of the usual behaviour of the overall system.  However, this technique can result in false alarm rate because it is difficult to find a correct model for general behaviour \cite{Dussel2010}. This approach can detect zero-day attacks\cite{LAHZA201848}. Yang et al. in \cite{Yang2005}, have proposed an anomaly detection using the Auto-Associative Kernel Regression(AAKR) with Statistical Probability Ration test (SPRT) and applied them to the network traffic. Goldenberg in \cite{GOLDENBERG201363}, have proposed a model-based IDS with the low false-positive rate for Modbus network. This approach is compassionate which also alarm anomalies, e.g., a message appearing out of order in the normal sequence. 
 Machine learning based techniques, i.e., probabilistic model-based technique, neural network-based approach, clustering model, multivariate based analysis are termed as statistical methods. Models are created based on these machine learning methods, and then these models serve as a reference model for intrusion detection. 
 \par
        Probabilistic model-based techniques are a data-driven unsupervised intrusion detection approach. Based on its analysis, it can distinguish between normal and critical states and removes the requirement for domain experts.
        The standard states are represented by a combination of the status and values that can be clustered into a finite group of dense clusters. The critical states take the form of noise, i.e., outliers,. It also extracts efficient detection rules from the identified states.     Almalawi et al. in \cite{Almalawi2016}, have also tested this algorithm on eight databases including five public databases.
        The presented algorithm approaches an average precision of 98\% in recognizing the critical states.
     \begin{table*}[h!]
    \centering
    \caption{Comparison of IDSs}
    \label{tab:Literature_survey_IDS}
    
    \begin{tabular}{p{1cm}p{3cm}p{3.5cm}p{3cm}p{3.5cm}}
        \noalign{\hrule height 2pt}
        \bf Source   &\bf Data type & \bf Input Data  & \bf\
        Technique &\bf Attacks handled
        \\\noalign{\hrule height 2pt}
        \cite{5168010} & Modbus/TCP simulated data & Time-stamp, Ip address & Anomaly & Probes, DoS, attempts to introduce rogue traffic.
        \\\hline
        \cite{Yang2005}  & SCADA simulated data& Link utilization, CPU usage, Login failure  &  Anomaly and Signature & DoS
        \\\hline
        
        \cite{Bigham2003}   & Simulated Data. & Extracted features  & Specification & Not mentioned
        \\\hline
        \cite{Cucurull}   & Real and Simulated Data & Extracted Features  & Anomaly & Draining attack, Grey hole attack
        \\\hline
        \cite{Rrushi2008}  & Not mentioned & Protocol data units (PDU) packets  & Anomaly & Not mentioned
        \\\hline
        \cite{DAntonio2006}   & Simulated Data. & Source and Destination IP, Source Port  & Behavior & DoS
        \\\noalign{\hrule height 2pt}
        
    \end{tabular}
\end{table*}   
\par Rrushi et al. in \cite{Rrushi2008}, tried to leverage the evolution of the content of the specific locations in random access memory into means of characterising the normalcy or abnormality of network traffic. The proposed algorithm uses estimation methods from probability theory and applied statistics to measure normal progressions of RAM content. 
  Yang et al. in \cite{Yang2014}, recommend a multilayer framework without undermining the availability of real-time data. The proposed algorithm analyses multiple attributes so that the provided solution can diminish various cyber-attack threats. Authors have also introduced a testbed to investigate simulated attacks. The proposal consists of 3 attributes, i.e., access-control whitelists, behaviour-based rules, and protocol-based whitelists. Access-control whitelists are the first list verified for allowing access.    
        Marton et al. in \cite{Marton2013}, proposed a combination of partial least squares (PLS) and principal component analysis (PCA)  to monitor abnormal behaviour.
        \par 
         Linda et al. in \cite{5178592}, presented an IDS based on neural network (IDS-NNM) model. The algorithm uses a union of two neural network algorithm i.e.Levenberg Marquardt and the error backpropagation. The IDS-NNM consist of two steps. In the first step, a particular training set is created. Later, the neural network starts training using that training set. Once the training set is generated, it is used in the network communication system to identify intrusion endeavours. 
\par \textcolor{black}{Kravchik et al. in \cite{Kravchik}, presented a study of detecting cyber attacks on ICS using convolutional neural networks on a Secure Water Treatment testbed (SWaT) dataset. Their research demonstrates that ID convolutional neural networks work better for anomaly detection as compared to other classification algorithms. Yang et al. in \cite{Yang2019} proposed a deep-learning-based network intrusion detection system for SCADA networks using the convolutional neural network (CNN) to identify the salient temporal patterns of SCADA traffic. They mainly used it to detect conventional and SCADA specific network-based attacks. }
\\ \\ \textbf { Specification based approach: } In specification-based approach, a model is constructed which imposes its predefined strategy and send an alert if the observed behaviour does not follow this policy. This technique defines what is allowable regarding patterns. It has the same purpose of the anomaly detection system. However, in specification-based approaches, a human expert define the policy for each specification manually. This approach causes a lower false positive rate due to a manually defined specification. Once the specification is set up, it can start functioning without the need for training.\\
    Goldenberg and Wool in \cite{GOLDENBERG201363}, discuss a specification based approach, for IDS which works for Modbus/ TCP networks. A fixed sequence of the query and the response is observed in Modbus traffic,    the fixed sequence is verified by operating over many SCADA network establishments. This DFA based IDS working on Modbus/TCP packets produces a very rigorous model, which has been evaluated using real traffic and it shows meagre false positive rate. 

Apart from this, behaviour patterns are associated with certain attacks. These type of attacks are used with the composition of other attacks. Moreover, some IDSs approaches have been proposed to specifically for resource constraint devices\cite{REEVES201274,IDS_resource_constraint,IDS_resource_constraint_2}.
Signature detection, anomaly detection, Machine learning based approaches are knowledge-based techniques. Behavioural detection approaches rely on the behaviour pattern. However, the specification approach uses knowledge as well as behavioural patterns.
Apart from the signature detection approach, all the procedures can detect new attacks. Anomaly detection, machine learning based methods, behavioural approaches matches the pattern while signature and specification based approach need predefine specifications.
Kraus et al. in \cite{Krau2016}, proposed a quick attack detection system. In the proposed ontology-based model, system logs provide suspicious logs. Suspicious logs with the previous vulnerability database lead to the detection of the ongoing attack. \par 
In \cite{Cuppens-Boulahia2008}, N. Cuppens and Boulahia presented an ontology that describes alert in IDMEF format. Based on the specific content of the fitting attack, an alert is generated and system use rules-based algorithm to react. Garcia et al. in \cite{Garcia2004}, has proposed a method to correlate alarm in a decentralised system. The proposed system can identify organised attacks of several attackers.
\par \textcolor{black}{Table \ref{tab:Literature_survey_IDS} shows a brief description of various intrusion detection systems, where data type represents whether the data used was simulated or collected on a real SCADA system  . Input data and attacks handled represent the input to the framework and the threat model for respective IDSs. The technique represents the IDS categorization.} Most of the IDSs do not specify the communication protocols. The data used for the verification of the systems are simulated due to the lack of the dataset. Traditionally, IDS is designed to detect a fixed type of attack, i.e., DOS, routing attack. There is an urgent need to develop an interface which can either combine various IDS and detect all possible attacks or develop an IDS which can detect various attacks. 
\par \textcolor{black}{ Belqruch et al. in \cite{Belqruch},  proposed to use Kippo, an SSH honeypot tool to log brute force attacks and shell interaction performed by attackers. Their aim is to distract the attacker from targeting the production server.} 
Researchers are creating testbeds for the SCADA systems to overcome the deficiency of well-validated datasets for the verification of IDSs. A brief survey of the testbeds is done in the next section.

\section{Testbeds for SCADA system}  \label{Testbeds}
Many approaches are used for the implementation of SCADA systems. We review some of them in the context of vulnerability assessment of SCADA protocols and system. The replication of a SCADA system can be physical, virtual-physical, virtual, hybrid. Table \ref{tab:testbed_list} shows a analysis of testbeds reported in literature. \textcolor{black}{Here, we studied different proposed testbeds based on the techniques used, the communication protocol used for simulating, the type of testbed and the replication strategy for the testbed. }

\begin{table*}[h!]
    \centering
    
    \caption{\textcolor{black}{Testbed list}}
    \label{tab:testbed_list}
    \begin{tabular}{p{5cm}p{4.5cm}p{2cm}p{2.75cm}}
        \noalign{\hrule height 2pt}
        \bf Testbed &\bf Technology used & \textcolor{black}{\bf Protocol } &\bf Type of Testbed
        \\\noalign{\hrule height 2pt}
        Cyber security backdrop \cite{Farooqui2014} & MATLAB/Simulink based tool utilizing TrueTime. & \textcolor{black}{ Modbus/TCP}  & virtual testbed 
        \\\hline
        TASSCS \cite{Mallouhi2011} & Opnet, PowerWorld Simulation System and Automatic Software Protection System (ASPS). & \textcolor{black}{ Modbus }& virtual testbed 
        \\\hline
        
        Testbed for cyber-power system setting \cite{Hong2011} & Power system simulation and sub-station automation based on Open Platform Communication (OPC) client/server architecture. & \textcolor{black}{DNP3 }& virtual testbed 
        \\\hline
        
        A HIL SCADA testbed \cite{Aghamolki2015a} & phasor Measurement Units (PMUs) synchronized with GPS reference signals & \textcolor{black}{ DNP3, Modbus}  & Hybrid model
        \\\hline
        
        A testbed for the gas-distribution system, water storage distribution and steel mining \cite{Morris2011} & Communication between HMI and UART-based MTU over 900MHz radio functioning as a repeater& \textcolor{black}{DNP3, Modbus} &Small-Scale Physical Models
        \\\hline

        A CPS testbed for Smart Grid \cite{ashok2011cyber} & Substations as two overcurrent relays each connected to a single Opal-RT RTDS (Real-Time Digital Simulator)
        &   \textcolor{black}{DNP3, IEC61850} & Small-Scale Physical Models
        \\
        \cline{2-2}
        & Two software based RTUs that communicate through TCP/IP to a control unit connected to HMI and historian (both working in active and redundant modes). & &
        \\\hline
        
        Industrial Control System (ICS) Security Testbed \cite{Korkmaz2016} & Electricity generation simulated using AC and DC motor pairs, PLCs and HMI. & \textcolor{black}{N/A} & Small-Scale Physical Models
        \\\hline
        
        Australian ICS security framework \cite{Foo2013} & Physical PLCs, VMware server, and networking hardware. & \textcolor{black}{N/A} & Small-Scale Physical Models
        \\\hline
        
        National SCADA Testbed \cite{U.S.DepartmentofEnergy2009} & 
        Bolstered by various labs supporting more than twelve test sites with full-size devices like a power grid. & \textcolor{black}{N/A} & Full-Scale Physical Models
        \\\hline
        
        MATLAB based testbed \cite{Chen2013} &MATLAB,Simulation
        packages such as OMNeT++, OPNET, and RINSE for the cyber layer simulation & \textcolor{black}{Modbus} & Hybrid Model
        \\\noalign{\hrule height 2pt}
    \end{tabular}
\end{table*}
Mallouhi et al. in \cite{Mallouhi2011}, have proposed a testbed using "PowerWorld simulation system", Opnet, and automatic "software protection system". Farooqui et al. in \cite{Farooqui2014}, have proposed a MATLAB based tool utilising TrueTime. The proposed Power Cyber test setup brings together VPN devices, relays to protect against overcurrent, Autotransformers, HMI and RTU software modules. 
Yang et al. in \cite{Yang2012_testbed}, proposed a testbed which contains SCADA software and communication infrastructure for investigating man-in-the-middle attack. 

\par A lot of theoretical security approaches have been presented in the last few years. However, the present research is still lacking practical approach \cite{Bou-harb2016}. Most of the security approaches need empirical data to train and test the system, In case of SCADA systems, these approaches fails due to lack of real empirical operational data. NIST had also suggested a set of guidelines in carrying out security assessment on SCADA systems.
However, the development of testbed is an expensive process. These type of event need substantial financial investments. Only some government-sponsored projects for testbed generation could afford such a vast investment \cite{Farooqui2014}. Moreover, the access to this testbed is restricted to the research community and academia. Thus, researchers focus on inexpensive and flexible approach for the development of SCADA test tools.
We have categorised SCADA testbeds into four categories based on replication strategy. 
\begin{enumerate}
    \item  \textbf{Physical Testbed : }
    A physical testbed corresponds to the replicating the existing SCADA system and industrial utility. Therefore, it demonstrates an excellent representation of the reliable, exact physical system. The scalability and cost of the physical system is a great issue due to the need for hardware stacks. The physical testbed can be further divided based on the scalability of the SCADA system, i.e., small-scale physical models and full-scale physical models. \par
    Industrial control systems security testbed is an example of a small-scale physical model. This model was based on power generation. The proposed system contains power generation units, real-time programmable logic controllers, drives and Human Machine interface. The presented testbed implements the process monitoring/ data collection. This data facilitates the different methods for exposing the cyber-attacks.
    \par The National SCADA Testbed (NSTB) \cite{U.S.DepartmentofEnergy2009} developed by United States Department of energy in Idaho National Labs(INL)is an example of full-scale physical models. It was designed for communication standards improvement. The maintenance of real hardware and software is a challenging task due to cyber-attacks. NSTB consists of a complex electrical grid with sixty-one miles, distribution lines of 13.8KV, transmission loop of 128KV and approximately three thousand points for monitoring. Many industrial protocols, e.g. internet protocol, GSM, ATM, MODBUS, DNP are supported in NSTB. Apart from the communication network, it supports firewall and VPN testing.  
 \begin{table*}[h!]
        \centering
         \caption{Testbed type analysis} 
        \label{tab:testbed_type_analysis}
        \begin{tabular}{p{3cm}p{4cm}p{5cm}p{3.5cm}}
            \noalign{\hrule height 2pt}
             \multicolumn{1}{c}{\textbf{Type of Testbed}} & \multicolumn{1}{c}{\textbf{Advantages}} & \multicolumn{1}{c}{\textbf{Disadvantages}}& \multicolumn{1}{c}{\textbf{Examples}}
            \\\noalign{\hrule height 2pt}
           Physical Testbed & 1.  Highest Degree of Fidelity.& 1. Difficult to reconfigure and sustain real hardware and software. & 1. The National SCADA Testbed \cite{U.S.DepartmentofEnergy2009}
        \\
       % \cline{2-3}
        & 2. Excellent reliability.& 2. Establishing a valid testbed is a costly operation.& 2. A testbed for SCADA security study, and pedagogy \cite{Morris2011}
        \\
       % \cline{3-4}
        && 3. Scalability is a big issue.& 3. A CPS testbed for smart grid \cite{ashok2011cyber}
        \\
       % \cline{3-4}
        && 4.  Poor repeatability.& 4. A proposed Australian industrial control system security curriculum \cite{Foo2013}
        %        \\
        %        \cline{4-4}
        %        && & National SCADA Testbed - Fact Sheet \cite{U.S.DepartmentofEnergy2009}
        %        

        \\\hline
        Virtual Testbed & 1. Secure from cyber-attacks as it enables a layer of abstraction between software and hardware.
        &1.  Incapable of reflecting the exact scenarios in the real SCADA systems due to the absence of real components and devices.& TASSCS \cite{Mallouhi2011}
        
        \\
       % \cline{2-3}
        & 2. Ease to develop and reconfigure.& 2. Low fidelity and reliability & VSSCADA \cite{VSSCADA}
        \\
      %  \cline{2-2}
        &3. Cost efficient and reliable & & SCADASim\cite{Qassim2017}
        \\
       % \cline{2-2}
        &    4. Good Scalability.&&

        \\\hline
        Physical-virtual testbed (Hardware-in-the-loop) & 1. Provide cost cutting measure for the design and testing of a wide variety of systems.
        & 1. Not cost efficient.&  A hardware-in-the-loop SCADA testbed \cite{Aghamolki2015a}.
        \\
       % \cline{2-3}
        & 2. The measurement data is more realistic.& 2. Scalability is a big issue.&
        \\
      %  \cline{2-2}
        &3. The communication pattern and latencies are more accurate.
        & &
        \\
      %  \cline{2-2}
        & 4. Vulnerability analysis and behaviour-based analysis are more feasible then simulated testbed.&&
        
        \\\hline
        
    %     Software simulated Testbed & 1. Ease to reconfigure, maintain, scalable.
    %     & 1. Low fidelity & \cite{Farooqui2014} 
    %     \\
    %   %  \cline{2-4}
    %     & 2. Cost efficient.& 2. Poor reliability and accuracy.& 2. An IDS testbed in a power system environment \cite{Hong2011}
    %     \\
    %     %\cline{3-3}
    %     &
    %     & 3. Incapable to capture the way computer network fail.&
        
    %     \\
    %   %  \cline{3-3}
    %     &&4. Incapable of reflecting the exact scenarios in the real SCADA systems due to the absence of real components and devices.&
        
    %     \\\hline
        
        Hybrid Testbed & 1. This approach enables the creation of a SCADA system using simulation, virtualization, emulation or simulation.
        & 1. Moderate cost-efficient.& 1. SCADA-SST \cite{Ghaleb2016}. 
        \\
      %  \cline{2-3}
        & 2. High degree of fidelity.& 2. Scalability is a big issue.& 
        
     \\\noalign{\hrule height 2pt}
        \end{tabular}
        
    \end{table*}
    \item  \textbf{Virtual Testbed : }
    The virtual testbed is developed to overcome the limitations of the software as well as physical testbed as it isolates activities in the test environment from the physical devices and the external components. It provides an abstraction between the software and hardware layer that provide an easy way to configure systems. Therefore, it is considered a controlled environment. TASSCS \cite{Mallouhi2011} falls in this category. It is developed by  NSF Center for Autonomic Computing, the University of Arizona.
     To simulate the control networks, e.g. Modbus, Allen-Bradley Data Highway, TASSCS uses Opnet tool and to simulate the operation behaviour; it uses PowerWorldi simulation system. Simulation of detection of attack/ protection is done using Autonomic Software Protection System (ASPS).
    \par 
    VSSCADA \cite{VSSCADA}, a  power system testbed,  virtualises all the hardware components maintain the actual behaviour of all the components. A testbed is purely software based on emulated communication allowing reconfiguribility of virtual systems for simullating much real control and monitoring  scenarios. VSSCADA supported Windows 7/Windows 8. It uses iFix, MatrikonOPC,  Power System Simulator for Engineering  (PSS/E) software to simulate HMI,  SCADA master control server, power system respectively.
 \par SCADASim framework \cite{Qassim2017} developed at the Royal Melbourne Institute of Technology, Australia is a software testbed.
  SCADASim uses OMNET++ to recreate SCADA components such as RTU, PLC, MTU, and communication protocols Modbus/TCP, DNP3. This can easily be scaled, integrated with other modules. It also proposes a concept of the gate, that is an interface between simulation and external environment. SCADASim supports multiple gates at the same time. It supports the denial of service, man-in-the-middle, eavesdropping, and spoofing attacks. \textcolor{black}{ Oyewumi et al. in \cite{Oyewumi} introduced the design of ISAAC testbed under development at the University of Idaho (ISAAC).  They designed ISAAC to be domain-independent, distributed, and reconfigurable. Alves et al. in \cite{Alves} proposed a modular, cost-efficient and portable testbed to replicate sophisticated SCADA Systems on a virtual simulation. They also demonstrated their approach by simulating real-world critical infrastructures. }
\item  \textbf{Virtual-physical Testbed : } It is also called Hardware-In-the-Loop(HIL) testbed. In this approach, the physical part or the entire critical infrastructure can be replaced by a computer model. HIL usually involves connecting control devices with control components and data acquisition. The measurement of HIL is more realistic and cost-effective. The measurements results of HIL, latencies, communication pattern, are more practical which reflect the data present in the actual control system. Apart from this, vulnerability analysis, as well as behaviour based monitoring, is realisable in HIL.
    An example of the virtual-physical testbed is presented in \cite{Aghamolki2015a} which is developed at USF Smart Grid Power System lab.
    The testbed was constructed to test energy management schemes, power grid cyber attack detection and prevention strategies. For visualisation, it uses PI-system.
    
    %\item  \textbf{Software Testbed : } Software based testbed is often a good alternative to the physical testbed. It employs computer software simulation through modelling the existing system to develop a software-based experimental environment. As compared to physical testbed, these are easy to reconfigure and maintain. However, these systems are more vulnerable to cyber-attacks. SCADASim framework \cite{Qassim2017} developed at the Royal Melbourne Institute of Technology, Australia is a software testbed.
     \item  \textbf{Hybrid Testbed : }
    In this approach, replication of the SCADA system is done using simulated, virtualised, emulated and physical devices. The main focus of Hybrid testbed is to provide a testbed for the cyber-security purpose. An example of the Hybrid testbed is explained in \cite{Chen2013}.
   
    In this testbed, a cyber-security scenario for Modbus worm attack was implemented. The architecture of Hybrid SCADA system is divided into two-layer architecture, i.e., hybrid cyber layer and virtual physical layer as shown. This two-layer can either be a real or simulated component.
    The architecture of the hybrid test system consists of sub-units: item condenser, a recycle compressor, two-stage reactor, vapour/fluid separator and product stripper. 
    \par
    Another example of hybrid SCADA testbed, i.e. SCADA-SST is presented in \cite{Ghaleb2016} for smart-grid and water tanks control. The proposed testbed is scalable, support hybrid scenarios, lightweight and can be widely used in different SCADA systems. It also supports malicious nodes templates, network attack scenarios.  It is specifically developed for  SCADA security evaluation and testing using OMNeT++ network simulator and  INET  framework. INET support the libraries needed to build communication network models.  SCADA-SST components behaviour is written in C++.  It also supports security analysis framework e.g.signature for the malicious node, attack scenarios, capturing and analysis of network traffic.
\end{enumerate}
Table \ref{tab:testbed_type_analysis} shows the advantages and disadvantages of the various categories of testbeds.
The physical testbed has the highest degree of fidelity but to maintain real hardware and software is a challenging task. It is also a costly operation.
Virtual testbeds have the lowest degree of fidelity and reliability. However, they are easy to develop. Therefore, various factors such as fidelity, reliability, cost, scalability issue should be considered to select the type of testbed. Now, hybrid testbeds are preferred because they have good accuracy and cost-effective. \par 
With the rapid advancement in technology, new technologies rapidly replace old techniques. In the next section, we will study the recent improvements, i.e., IoT based  SCADA system.

\section {Recent advances in SCADA} \label{Recent advances in SCADA}
The future Internet is considered as another game-changing idea for traditional SCADA frameworks. The current SCADA frameworks use a combination of characteristics of old and new features, due to which, their security is in greater danger. Generally, the SCADA system is inflexible, static and follows centralised architecture. These weakness limit the SCADA system interoperability. Therefore, to overcome the limitation of the existing SCADA, a sensor cloud-based SCADA infrastructure has been proposed. We can say that the integrated SCADA systems, an amalgamation of industrial business systems and the IoT,  is more prone to attacks in comparison to the traditional SCADA due to the larger exposed space \cite{Sajid2016}. Wei Ye and John Heidemann in \cite{Ye2006} introduced a new cloud-based framework which is capable of virtualising a wide range of sensing frameworks, comply new techniques for data processing, use cloud computing for managing a large amount of data collected from sensors.\par 
Alcaraz et al. in \cite{Alcaraz2011}, proposed VS-Cloud, a virtual SCADA architecture. His main focus is cloud storage. The SCADA system should offer dynamic sensing services management, i.e., It should allow dynamic creation and configuration of the offered services. The privacy of data should be provided. The proposed system should be scalable, fault tolerant, inter-operable, and energy-aware  \cite{BenDhief:2016:NSC:3007120.3007169}. \par 
% To differentiate wireless sensor networks, IoT, and CPS under IOT architecture is a difficult task. Fig. \ref{fig: WSN overlapping} provide a general CPS model which shows overlapping concepts.

% \begin{figure}[h!]
%     \centering
%     \includegraphics[width=3.5in]{overlap_color1}
%     \caption{CPS, IOT, WSN overlapping }
%     \label{fig: WSN overlapping}
% \end{figure}
IoT provides interconnectivity among various real-time sensors and other intelligent electronic devices. A typical IoT application platform is used for data analysis, SCADA PLC, queries, and reporting, remote terminal, process control, the web, cellular App, Historian, monitoring. Therefore, it has become a tremendous development in the area of real-time industrial infrastructure.
\par
Industrial IoT is a new revolution in smart industrial sectors that provide enhanced automation and information sharing facilities manufacturing. It is a combination of cloud computing, cyber system, and connectivity. A smart industrial system based on IoT system can predict the failure cases using the network devices.
\par
Moreover, industry system is searching for solutions that can provide fault tolerance, scalability, availability, and flexibility. One proposed solution is to integrate the CPS with IoT using cloud computing services. However, traditional SCADA systems cannot properly measure security parameters. The integration of traditional SCADA systems with IoT is more vulnerable to security threats. Therefore, these future concepts need more research efforts \cite{Sajid2016}. %, \cite{Spinal-epidural-general1990abibid}.

Real-time monitoring, Pay-per-use, licenses, cheaper capital and operating expense are the advantages offered by cloud-service \cite{Iosup2014}. Cloud-service providers handle the maintenance, upgrade of these systems. Once they are upgraded, they are available to all users instantly. The main concern of cloud-SCADA is security and performance issues \cite{Daalder}. 
\par 
Tracking of hackers, information leakage, latency time, privacy issue \cite{Sajid2016} reliability of the cloud servers should also be taken into consideration before shifting to cloud-SCADA. \cite{whitepaper_cloud_computing_SCADA, Church2017}.  The communication link, relying on cloud-based communication, can suffer from the Man-in-the-Middle attack, DoS attacks because the adversaries can still sniff, alter, or spoof the information on the network. The reliance on cloud communication opens more back-doors to the SCADA systems and critical infrastructure. The security risks in the traditional system will be carried forward owing to the communication protocols used like Modbus/TCP, IEC 40, and DNP3 which are suffering from lack of protection.
Moreover, these systems use commercial off-the-shelf solution rather than the proprietary solution. The information communicated to the cloud is divided locally. The probe of system application running on the cloud can be done, and therefore, these can be attacked by the attacker. 

\section{Future Research Directions} \label{Future Research Directions}
Even with the advanced security algorithm, a lot of attacks on SCADA system have been detected. This section highlights the future research scope abridging the gap between the current state of SCADA and an advanced, robust SCADA system.
\begin{enumerate}
    \item \textbf{Attacks Database :}
    The database of the security incidents is required to analyse the various dimensions of attacks to develop strategies to prevent similar attacks in future. \textcolor{black}{Datasets KDD99 \cite{Tavallae}, NSS-KDD \cite{zgr2016ARO}, DARPA  \cite{DARPA} are outdated and not synchronised with modern SCADA architecture. NVD dataset \cite{nvd_database} contains common vulnerabilities in all domains that fail to focus on SCADA specific vulnerabilities. There is no update to RISI database \cite{Risidata} since 2015.} Therefore, there is no proper database which has covered all security incidents. One global repository for all these incident should be built. This repository should be publicly available to researchers to analyse these attacks. Industries should also report the attacks on their system rather than hiding it to save their image. Then only zero-day attacks can be handled.
    \item \textbf{Scalable Testbeds \& validation techniques: }
    In the literature survey of testbed physical, virtual, virtual-physical, software, and hybrid testbed have been studied. The development of testbed is a costly process which needs a huge amount of funding. There is no such testbed which is cost-efficient, scalable, and have a high degree of fidelity. The researcher should focus on the scalable, higher degree of fidelity, cost-efficient, and interpolation solution. New communication protocols, new risk-assessments techniques as well as IDS need to be validated before deploying directly to the field. There is an urgent requirement for trust-worthy validation approaches to assess the reliability of new techniques for the safety and security of SCADA systems \cite{Lyu}.
    
%     \item \textbf{Improved Risk assessment techniques}
%     Many risk assessments approaches, frameworks have been proposed. The survey of risk assessment techniques gives an indication of the scope of further research in this area. The present approaches can be enhanced for zero-day attacks. Cherdantseva et al. in \cite{Cherdantseva2016} indicate the requirement for a general method which can incorporate all aspects of the risk management process.
    
    \item \textbf{ IDSs for SCADA :}
    Mitchell in \cite{Mitchell2014} suggests more research is required to define the performance metrics for the validation of IDSs. In most of the analysis, only attack discovery rate, false positive and false negative rates are provided. Time taken to detect the attack, an essential standard for performance measurement is a missing parameter from current research. Therefore, even if it is guaranteed that IDSs will detect the attack and the latency is high, the attacker will have sufficient time to damage the system. We did not find any paper which compares the IDSs based on the placement of the detection system. Moreover, research work focuses on the development of a detection system for specific attack types, i.e., routing and DOS attack. Different attack detection schemes which are running under similar operational settings can be evaluated in further research.\par
    Significant work has been done in the knowledge-based IDSs. However, these systems are still not capable of handling zero-day attacks. It is a challenging task to define acceptable behaviours upon environment change. The knowledge-based IDSs are not reliable for unknown attacks. The behaviour of each attack differ from others, so researcher should focus on identifying the attack model. Therefore researchers should make more effort to further refining the threshold monitoring techniques. Nivethan et al. in \cite{Nivethan2016}, propose a model for Dynamic rule generation for SCADA intrusion detection. These threshold model should be dynamic which learns as per the severity of past attacks. The priority for IDSs should be eviction in case the attacker is persistent, repairing in case the attacker is transient, establishing attribution for the attack in case the attacker is ineffective. 
    \par
    Zhu et al. in \cite{Zhu2017}, propose that SCADA system security must be an overlap of computer security, communication network, and control engineering. IDSs should be able to record the features of a specific SCADA system, i.e., the versatility of the physical system, communication pattern, system architecture, etc. A new area of research can be alert post processing for reducing false positive alert, as well as the development of techniques for alert correlation. Multi-step intrusions techniques can be used to correlate isolated intrusions \cite{WANG20062917}. 
    \par 
Moreover, not all intrusions can be prevented, use of honeypots, honeynets, etc. is an attractive approach \cite{Belqruch}. Shakarian et al. in \cite{Shakarian7509471}, proposed a new and realistic approach to delay the impact of intrusion in-spite of stopping it. This will help to minimise the probability that the intrusion reached its goal by giving the target system more response time. These kinds of techniques integrated with SCADA IDSs can help to avoid catastrophic events. 
    \item \textbf{New communication protocol :}
    In communication protocols, the focus is needed on the application and network layer security. Network security protocols should be integrated into these communication protocols.  Communication protocol for IoT-cloud based SCADA, i.e., a reliable, secure, scalable, open, low latency communication protocol is the new focus for the researcher. With Industry 4.0 evolution, IoT protocols are used in the SCADA system. These protocols lack reliability, raising the need for reliable communication protocols.
    Rezai in \cite{Rezai2016}, raised the issue of an efficient key management scheme. In the case of SCADA systems, network cryptographic solutions are not sufficient in blocking the attacks. There is still a need for extensive research for more robust cryptographic solutions, in-protocol authentication techniques, distributed security mechanisms which apply to SCADA systems. 
    \item \textbf{Safe and Secure architecture and operating system : }
     DOS, VMS, and UNIX operating systems, which have various vulnerabilities, were mostly used in SCADA. Now, Linux and Microsoft Windows-based operating systems have displaced DOS with UNIX based SCADA. However, Linux and Windows suffer from their vulnerabilities due to the large source code for operating systems. Microkernel architecture based operating systems can be used to reduce the attack surface for SCADA systems \cite{hentea2008improving}.  Apart from security, Safety guidelines should always be followed to the maximum extent to avoid acceptable risks. SCADA systems can be secured by utilising a more error-resistant architecture, secure and robust operating system, and usage of secure programming languages. Recently, Kaspersky launched a secure operating system for SCADA which does not have traces of Linux \cite{ComingkasperkeyOS,kasperkeyOS}. Additionally, secure architectures for SCADA have been proposed recently\cite{SecureSCADAArchitecture, SecureArchitecture}. Safety of CI is an important concern. Safety protocols need to be mandatory. With IoTization, the safety of the end-devices is a big concern as these cheap devices are from different vendors which rarely follow safety guidelines.  
     
    \item \textbf{Research focus for IoT-Cloud SCADA :}
    Sajid et al. in \cite{Sajid2016} give some research proposals for a secure IoT-cloud based SCADA system. Integration of IoT-cloud in traditional SCADA system offers new vulnerabilities and opportunities to share data/information/services over the web \cite{Fazio2012}.
    There is a dire need to grow new strategies that are fit for managing complex and large-scale frameworks. Research should be focused on continuously enhancing the security of these systems. In IoT-cloud based system, bandwidth overload and latency are a big issue. These parameters are dependent on cloud service providers. Delay in decision making, i.e., latency delay can cause a loss of production. So research should be focused to make this system robust. The high bandwidth and low latency providers should be encouraged. The potential of these systems is dependent on the cooperation among the responsible platforms. 
    \par  To assure industries about these complex collaboration, more research is required. New development tools which can handle the complexity of service creation are needed. Apart from these, more productive and upgraded use of worldwide assets is needed. Sustainable development goals should also be considered in parallel to achieve robustness, scalability, reliability, real-time system. In IoT based system, a massive amount of data gets generated. Therefore, the security, analytics, storage, and complexity of this data are the principal concern. 
 
\end{enumerate}

\section{Conclusion}
 SCADA systems have evolved into sophisticated complex open systems based on advanced technology systems connected to the Internet. This medium has lead the SCADA system vulnerable to attackers. Over the period, many attacks on SCADA industrial control system have been reported. The impact and severity of these attacks varied. The smooth and genuine operation of SCADA framework is one of the key concern for the enterprises, because the consequences of break down of SCADA system may range from financial misfortune to natural harm to loss of human life.
\par As per the analysis of RISI SCADA attack database, the count of attacks seems quite less. There are two aspects to this analysis. First, most of the industries do not report cyber-attacks on their control system or SCADA for the sake of their reputation. Second, the impact of these attack can vary up to loss of human life.
Therefore, there is an urgency for securing SCADA systems. 
% A risk assessment of the SCADA system can provide an overview of such vulnerabilities.
However, there is a requirement of a responsive intrusion detection system which can alert the system managers about the possible attack on the system and network. These detection systems can use signature,  specification, behaviour or machine learning based models for the enhanced security. There are many cryptographic approaches discussed in the research community, but we kept it out of scope for our review, as modern SCADA include a lot of resource constraint devices, which render the cryptographic solutions inappropriate.
\par This paper gives a structured and multi-dimension overview of security of SCADA systems. The major contributions of this paper are:-
\begin{enumerate}
\item It provides a novel approach to SCADA security by linking various security aspects.
\item A comprehensive analysis of RISI attack database, Intrusion Detection Systems,
% Risk assessment approaches
and SCADA testbeds.
\item  Due to the IoTization and cloud-based SCADA, the research problems for secure SCADA has been widened. Therefore, a brief discussion about future research directions is done in section \ref{Future Research Directions}.
\end{enumerate}
This review indicates that despite a lot of approaches present for IDS, 
% risk assessment, 
and testbeds, there is still a lot of scope for further improvements.
IDSs can be improved in sectors of placement policy, validation strategy, attack coverage, low latency and low false positive rate. 
% Risk assessments approaches can be researched further concerning validation, software toolkits, the reliability of data, etc.
Similarly, testbeds can be improved pertaining to cost, scalability and high fidelity solution. \par
Apart from this, currently industries are shifting to cloud-based SCADA systems as they are economical and easily scalable. But cloud-based SCADA system is hampered by performance issues, i.e., high latency and low bandwidth. Therefore, there is a need to build a viable and efficient system architectures and frameworks to model such issues.
%Nothing should interrupt the references.
% \bibliographystyle{iet}           % Style BST file.

\bibliographystyle{unsrt}  
\bibliography{ref}

\end{document}